\documentstyle[aps,prd,eqsecnum,twocolumn,epsf]{revtex}

\newcommand{\beq}{\begin{equation}}
\newcommand{\beqn}{\begin{eqnarray}}
\newcommand{\eeq}{\end{equation}}
\newcommand{\eeqn}{\end{eqnarray}}

\newcommand{\gsim}{\mbox{\raisebox{-1.ex}{$\stackrel
       {\textstyle>}{\textstyle\sim}$}}}
\newcommand{\lsim}{\mbox{\raisebox{-1.ex}{$\stackrel
       {\textstyle<}{\textstyle \sim}$}}}
\newcommand{\square}{\kern1pt\vbox{\hrule height
1.2pt\hbox{\vrule width 1.2pt\hskip 3pt
     \vbox{\vskip 6pt}\hskip 3pt\vrule width 0.6pt}\hrule
height 0.6pt}\kern1pt}

\begin{document}

\draft
\twocolumn[\hsize\textwidth\columnwidth\hsize\csname
@twocolumnfalse\endcsname

\title{Quintessence in {\bf a} Brane World}
\author{Shuntaro Mizuno$^1$ and Kei-ichi Maeda$^{1,2}$ \\~}
\address{$^1$Department of Physics, Waseda University, Okubo 3-4-1,
Shinjuku, Tokyo 169-8555, Japan\\[-1em]~}
\address{$^2$ Advanced Research Institute for Science and Engineering,
Waseda University, Shinjuku, Tokyo 169-8555, Japan\\[-.7em]~}

\date{\today}
\maketitle
\begin{abstract}
We reanalyze a new quintessence scenario in
a  brane world model, assuming that a quintessence
scalar field is confined in our 3-dimensional brane world. We study
three typical quintessence models : (1) an inverse-power-law potential,
(2) an exponential potential, and (3) kinetic-term quintessence
($k$-essence) model. With an inverse
power law potential model ($V(\phi) =
\mu ^{\alpha + 4} \phi^{-
\alpha}$), we show that in the quadratic dominant stage, the
density parameter of a scalar field $\Omega_\phi$ decreases as
$a^{-4(\alpha-2)/(\alpha+2)}$ for $2<\alpha < 6$, which is  followed by
the conventional quintessence scenario.
This
feature provides us wider  initial conditions for a successful quintessence.
In fact, even if the universe is initially in a scalar-field dominant,
it eventually evolves into a radiation dominant era in the
$\rho^2$-dominant stage.

  Assuming  an equipartition condition,  we discuss constraints
on parameters, resulting that $\alpha\geq 4$ is required. This constraint also
restricts the value of the 5-dimensional Planck mass, e.g.  $4 \times 
10^{-14}m_4 \lsim  m_5 \lsim  3 \times
10^{-13}m_4$ for
$\alpha=5$.
For an exponential potential model
$V=\mu^4\exp(-\lambda \phi/m_4)$, we may not find a natural and  successful
quintessence scenario as it is.
While, for a kinetic-term quintessence, we find  a tracking solution
even in $\rho^2$-dominant stage,
rather than the $\Omega_\phi$-decreasing solution for  an inverse-power-law
potential. Then we do find a little advantage in a brane world.
Only the density parameter increases more slowly in the
$\rho^2$-dominant stage, which provides a wider initial condition for a
successful quintessence.
\end{abstract}
\vskip 1pc 
\vskip 1pc
]

\section{Introduction}                           %

Recent observation  of the angular   spectrum of  the
cosmic microwave background (CMB) in a wide range of $\ell$,
besides the results for small $\ell$ obtained by COBE, suggests
  a flat universe\cite{CMB}. By combining this result with the
measurements of high-redshift supernovae\cite{S-novae}, which predicts
  that the expansion of the universe is accelerating now,
we are forced to
recognize  the existence of a cosmological constant, or a kind of
dark energy, which value is  almost  the same order of
magnitude as the present mass density of the universe. From the viewpoint  of
particle physics, however, it is quite difficult to explain such a tiny value
by
$14
$ orders of magnitude smaller than the  electroweak scale. The
failure of theoretical  explanations for  the present value of the
cosmological constant is known as  the so-called ''Cosmological Constant
Problem."
\cite{C. Problem}

Since there is no plausible theoretical idea
to explain such a small cosmological constant, it
may be more plausible that such a tiny value is achieved through the dynamics
of a fundamental field. This idea is called a decaying cosmological
constant\cite{D-C-C},\cite{Overduin}.
Among such models, the so-called quintessence seems to be
more promising.
It shows
an interesting property, i.e. a quintessence field follows a
common evolutionary track just as an attractor in a   wide  range of the
initial conditions, so that the cosmology is extremely insensitive to the
initial conditions \cite {Quintessence},\cite {Ratra-Peebles}, \cite{tracker solutions},
\cite{scaling solutions}. Some of them (so-called "tracker-fields") may avoid
the coincidence problem. According to a successful quintessence scenario,
the energy of a scalar  field tracks the radiation energy (or matter
energy) for a rather long time in order not to affect the nucleosynthesis at the
radiation dominant era, and the structure formation at the matter dominant
era, and then becomes dominant just before the present time.

There
seems  to  be,
however, still a kind of fine-tuning problem  for these
models.
Although the initial energy of a scalar field could be the same as that of
radiation fluid, most contribution is from its kinetic energy.
In the kinetic dominant case, the scalar field behaves as a massless field,
which is equivalent to a stiff matter. Hence its energy density drops as
$a^{-6}$ which is much faster than the radiation energy.
Then the kinetic term drops in the radiation dominant universe until a 
tracking
solution is found. This is why the density parameter of a scalar field
decreases before reaching a tracking solution.
Since we have to tune the mass scale of a potential for a successful
quintessence, the potential term cannot be so large.
Hence, an equipartition condition, which may be expected in the early stage of
the universe,  seems to be unlikely in such a model.
  Another unsatisfactory
point is that if the universe starts in a scalar-field dominant condition, the
radiation dominant universe is never recovered.
Then we are not able to  find the present universe.
  Some modified models within the conventional
gravity  theory have been proposed to solve such problems
\cite{Grad},\cite{Ex-exponential}.

As for the early stage of the universe, recently we have a
new interesting
idea, which is called brane cosmology. In a brane world scenario,  our
universe is embedded in higher dimensions and  standard-model  particles are
confined to four-dimensional hypersurfaces (3-branes), while gravity is
propagating in higher-dimensions (a bulk) \cite{String} -\cite{3-brane}.
Among them Randall-Sundrum's second model is very interesting because
it provides a new type of compactification of gravity.
  Assuming the 3-brane has positive tension and is
embedded in 5-dimensional anti de-Sitter bulk spacetime, the conventional
four-dimensional gravity theory is recovered, even though the extra
dimension is not compact \cite{RS}. While, in a high energy region,  gravity
theory is very different from the four dimensional
Einstein theory\cite{SMS}. Many authors  discussed the geometrical aspects
and its dynamics (For a review, see, \cite{Brev}),  as well as
cosmology
\cite{Bcmgy1}, \cite{Bcmgy2}, \cite{Bcmgy3}.

The main modified point from conventional cosmology is the appearance
of the  quadratic term of energy-momentum and dark radiation.
Since the quadratic term changes the dynamics of the universe in its early
stage,  we  expect some improvement for a quintessence scenario. In
\cite{maeda}, it is shown that the quadratic term indeed drastically
changes the evolution of a scalar field with an inverse-power-law 
potential and
then  the density parameter of a scalar field  decreases  in time until 
conventional cosmology is recovered.
  This result helps to
construct a more natural and successful quintessence scenario.

The purpose of the present paper is to present a full analysis of the
previous letter, including a numerical analysis, and to
study other types of quintessence models in the brane world scenario.
In
Sec.\ref{sec.2}, we present the basic equations assuming  the
Randall-Sundrum II model. We then
study three typical quintessence models,  an inverse-power-law potential model
(in Sec.\ref{sec.3}), an exponential potential model  (in Sec.\ref{sec.4}), and
a kinetic term quintessence ($k$-essence) model (in Sec.\ref{sec.5}). Sec.
\ref{sec.6} is devoted to the conclusions.

   \section{BASIC EQUATIONS }
   \label{sec.2}
We start with the Randall-Sundrum type II brane scenario\cite{RS},
because the model is simple and concrete. It is,  however, worthwhile noting
that the present mechanism may also work in other types of brane world
models, in which a quadratic term of energy-momentum tensor generically
appears.
In the brane world,  all matter fields and forces except gravity
are confined on the 3-brane in a higher-dimensional spacetime.
As for gravity, in contrast to the other type of brane scenario, in the
    Randall-Sundrum type II model, the
extra-dimension is not compactified, but gravity is confined  in the
brane, showing the Newtonian gravity in our world.
Since the gravity is confined in the brane, it can be described by the
intrinsic metric of the brane spacetime. By use of Israel's thin shell
formalism and assuming
$Z_2$  symmetry, the gravitational equations on the 3-brane is
given by
\beqn
^{(4)}G_{\mu\nu}=-^{(4)}\Lambda g_{\mu\nu}+\kappa_4^2
T_{\mu\nu}+\kappa_5^4 \pi_{\mu\nu}-E_{\mu\nu},
\label{geq}
\eeqn
where $^{(4)}G_{\mu\nu}$ is the Einstein tensor with respect
to the intrinsic metric $g_{\mu\nu}$, $^{(4)}\Lambda$ is the 4-dimensional
   cosmological constant, $T_{\mu\nu}$
represents the energy-momentum tensor of matter fields confined on the
brane  and
$\pi_{\mu\nu}$ is quadratic in
$T_{\mu\nu}$\cite{SMS}. $E_{\mu\nu}$ is a part  of the 5-dimensional
Weyl tensor and carries some information about bulk geometry.
$\kappa_4^{\;2}=8\pi G_4$ and
$\kappa_5^{\;2}=8\pi G_5$ are 4-dimensional and 5-dimensional
gravitational constants,  respectively. In what follows, we use the
4-dimensional Planck mass $m_4 \equiv \kappa_4^{~-1}
=(2.4\times10^{18}{\rm GeV})$ and the 5-dimensional 
Planck mass $m_5\equiv
\kappa_5^{~-2/3}$, which could be much smaller than  $m_4$.
We also assume that $^{(4)}\Lambda $ vanishes.

Assuming the Friedmann-Robertson-Walker
spacetime in our brane world,  we find the Friedmann equations
from Eq.(\ref{geq}) as
\beqn
H^2+{k\over
a^2} &=&\frac{1}{3m_4^{\;2}}\rho+\frac{1}{36m_5^{\;6}}\rho^2+\frac{{\cal
C}}{a^4}
\label{fr1}
\\
\dot{H}-{k\over a^2}&=&-\frac{1}{2m_4^{\;2}}\left(P+\rho\right)
-\frac{1}{12m_5^{\;6}}\rho\left(P+\rho\right)-\frac{2{\cal C}}{a^4}
\label{fr2}
\eeqn
where $a$ is a scale factor of the Universe, $H=\dot{a}/a$ is its Hubble
parameter, $k$ is a curvature constant,
$P$ and
$\rho$ are the total pressure and total energy density of matter fields.
${\cal C}$ is a constant,  which describes "dark"    radiation
coming from $E_{\mu\nu}$
\cite{Bcmgy1}. In what follows, we consider only the flat Friedmann model
($k=0$) for simplicity.

As for matter fields on the brane, we consider   a scalar field $\phi$
as well as the conventional radiation and matter fluids, i.e.
$\rho=\rho_{\phi}+\rho_{\rm r}+\rho_{\rm m}$, where $\rho_{\phi},
\rho_{\rm r}$ and
$\rho_{\rm m}$ are the energy densities of scalar field $\phi$,
of radiation fluid and of  matter one, respectively.
Although we can consider a 5-dimensional scalar field living in the
bulk\cite{maeda_wands}, we shall focus only on a 4-dimensional scalar
field confined on the brane.
The origin of such a scalar field might be a condensation of some
fermions confined on the brane.

Since the energy of each field on the brane is conserved in the
present model, we find the dynamical equation for such a 4-dimensional
scalar field as a conventional one, i.e.
\beqn
\ddot{\phi}+3H\dot{\phi}+\frac{dV}{d\phi}=0,
\label{sde}
\eeqn
where $V$ is a potential of the scalar field.
The energy density of
the scalar field is
\beqn
\rho_\phi=\frac{1}{2}\dot{\phi}^2 + V(\phi),
\label{sed}
\eeqn
and for the radiation and  matter fluids, we have
\beqn
\dot{\rho_{\rm r}}+4H\rho_{\rm r}&=&0,
\label{ecr}
\\
\dot{\rho_{\rm m}}+3H\rho_{\rm m}&=&0 .
\label{ecm}
\eeqn

As the Universe expands, the energy density  decreases.
This means that the quadratic term was very important in the early stage
of the Universe. Comparing two terms (the conventional energy density
term and the quadratic one), we find that the quadratic term dominates
when
\beq
\rho > \rho_c \equiv 12 m_5^{~6}/m_4^{~2} .
\eeq
When the quadratic term is dominant,
the expansion law of the Universe is modified.
For example, in  the radiation dominant era, i.e. $\rho_{\rm
r}\gg\rho_{\rm m}, \rho_\phi$, the Universe expands as
$a\propto t^{1/4}$, in contrast with $t^{1/2}$ in the conventional
radiation dominant era.

Since we are interested in a quintessence scenario and then the dynamical
behavior of a scalar field, we shall calculate the density parameter of the
scalar field, which is a   ratio of the energy density of the scalar
field to that of total energy density ($\Omega_\phi$).
Ignoring dark radiation, we find
\beqn
\dot{\Omega_\phi}=\frac{H\rho_{\rm
r}}{\rho^2}\left(4V(\phi)-\dot{\phi}^2\right).
\label{omg}
\eeqn
When the quadratic term dominates in the early stage, the Hubble
expansion rate decreases. As a result, the friction term in Eq.
(\ref{sde}) becomes small and then the dynamics of the scalar field will
drastically change from the conventional one even for the same potential.
It is easily expected  that a kinetic term will play a more important role
in the energy density of a scalar field.
Consequently, if $\dot{\phi}^2 > 4V(\phi)$, the density parameter of
a scalar field will decrease in time.
In particular, this feature may be very important in a quintessence
scenario, because most quintessence models assume very small
potential energy, initially, for a successful scenario.
We will show that this interesting feature of a scalar field
is found in some quintessence models and the initial conditions for a
successful scenario become much wider.

In the following sections, assuming that a quintessence field  exists
in a quadratic term dominant era, we investigate the dynamical behavior
of a quintessence scalar field and search for  a
natural and successful quintessence scenario.
We study three types of quintessence fields: models with an inverse power
law potential, with an exponential potential and a kinetically driven
model.

   \section{Inverse power law potential}
\label{sec.3}
In this section, we  investigate a model with an inverse power law
potential\cite{Quintessence},\cite{Ratra-Peebles}, i.e.
\beq
V\left(\phi \right)  =
\mu^{\alpha+4}\phi^{-\alpha},
\label{invpowpot}\eeq
where $\mu$ is a typical mass scale of the potential, which is not fixed
here. Although this potential is not renormalizable, it may appear as an
effective potential for a kind of  fermion
condensation in a   supersymmetric QCD model\cite{ivplpotentilorigin}.
The $SU(N_c)$ gauge symmetry is broken by a pair condensation  of
$N_f$-flavor quarks, and  the
effective potential for a fermion condensate field $\phi$ is given by
(\ref{invpowpot}) with
$\alpha=2(N_c+N_f)/(N_c-N_f)$.

The quintessence
scenario with this potential has been well studied  in the conventional
universe\cite{tracker solutions}. We therefore turn
our attention mainly to the
quadratic term ($\rho^2$) dominant stage.
Although some analytic solutions are already given in Ref.\cite{maeda},
we  first summarize those and then show numerical analysis.   We then
discuss a natural  quintessence scenario by this model and give some
constraints for unknown parameters such as $m_5$ .

   \subsection{Analytic solutions in $\rho^2$-dominant stage.}
In the conventional universe,  matter energy density is negligibly
small in comparison with   radiation
energy density in the early stage of the universe.
Then we know that it is also  the case in the $\rho^2$-dominant stage.
Hence we consider only  radiation fluid and a scalar  field $\phi$.
As for the initial conditions, we have two possible cases:  one is
the radiation dominant case  and the other is the $\phi$-dominant one.
We discuss these separately.

\vspace*{0.3cm}
$\left(1\right)$ radiation dominant initial conditions
\vspace*{0.3cm}

In this case, the scale factor expands as $ a \propto t^{1/4}$
and the equation for the scalar field
(\ref{sde}) is now
\beqn
\ddot{\phi}+\frac{3}{4t}\dot{\phi}-\alpha\mu^{\alpha + 4}\phi^{-\alpha-1}
=0.
\label{ivsde}
\eeqn

   We find an analytic solution for $\alpha<6$\cite{maeda}, that is
\beqn
\phi &=&\phi_0 \left( {t\over t_0}\right)^{\frac{2}{\alpha + 2}},
\label{sevkpc}\\
&& {\rm with}~~~
\phi_0^{\alpha+2}={2\alpha(\alpha+2)^2 \over 6-\alpha}\mu^{\alpha+4}t_0^2
\label{phi0}
\eeqn
where $t_0$ and $\phi_0$ are integration constants.
The constraint (\ref{phi0}) requires $\alpha<6$.
The energy density of the scalar field evolves as
\beqn
\rho_{\phi} &=&
{3\alpha(\alpha+2) \over
6-\alpha}V_0\left({t\over t_0}\right)^{-\frac{2\alpha}{\alpha + 2}}
\nonumber\\
& =& {3\alpha(\alpha+2) \over
6-\alpha}V_0\left({a\over a_0}\right)^{-\frac{8\alpha}{\alpha + 2}},
\label{sfedqrrde}
\eeqn
where $V_0=V(\phi_0)$ and $a_0=a(t_0)$.
As for the density parameter $\Omega_\phi$,
we find that
\beq
\Omega_\phi = {\rho_\phi \over \rho_{\rm r} +\rho_\phi} \approx
{\rho_\phi \over \rho_{\rm r}} = \Omega_\phi^{(0)} \left({a\over
a_0}\right)^{\frac{4(2-\alpha)}{\alpha + 2}},
\eeq
where $ \Omega_\phi^{(0)}={3\alpha(\alpha+2) \over
6-\alpha}V_0/\rho_{\rm r}(t_0) $
because  $\rho_{\rm r} \propto a^{-4}$.

If
$
\alpha > 2$, just contrary to the tracking solution,  the scalar field
energy decreases faster than that of the radiation.
This result is confirmed by Eq.(\ref{omg}) with the fact that  the
kinetic energy of a scalar field
$\rho_\phi^{~(K)}=\dot{\phi}^2/2$ turns out to be larger than $4V$ as
$\rho_\phi^{~(K)}/(4V)=2\alpha/(6-\alpha)$.

If $ \alpha = 2$,  the
scalar field energy drops at the same rate as that of the radiation until
the conventional universe is recovered. This is the so-called "scaling"
solution.
If $ \alpha < 2$, the scalar field energy
decreases slower than the radiation energy,  and eventually the scalar
field dominates the radiation.

What happens for $\alpha \geq 6$ ?
Since there is  no asymptotic solution for which a kinetic term
balances with a potential term, we expect that a kinetic term
dominant solution is obtained asymptotically.
It is easy to show that a potential dominant (or slow rolling) condition
is not preserved asymptotically.

   Assuming that a kinetic term is dominant, Eq.
(\ref{ivsde}) is now
\beqn
\ddot{\phi}+\frac{3}{4t}\dot{\phi}=0,
\label{ivsdekd}
\eeqn
finding an asymptotic solution
\beqn
\phi \propto t^{\frac{1}{4}}.
\label{sevkdc}
\eeqn

With this solution, we find that
\beqn
\rho_{\phi}^{~(K)} &=& {1\over 2}\dot{\phi}^2 \propto t^{-3/2}
\\
\rho_{\phi}^{~(P)} &=&\mu^{\alpha+4} \phi ^{-\alpha} \propto
t^{-\alpha/4} ,
\eeqn
    where $\rho_{\phi}^{~(K)}$ and $\rho_{\phi}^{~(P)}$ denote the kinetic
   and the potential terms of the energy density of a scalar field,
respectively. Hence, if $\alpha \geq 6$, kinetic term dominance is
guaranteed.  This is also confirmed by numerical calculation.

\vspace*{0.3cm}
$\left(2\right)$ $\phi$-dominant initial conditions
\vspace*{0.3cm}

If a scalar field initially dominates radiation, what we will find in
the dynamics of the scalar field?
In the conventional universe, once a quintessence field dominates,
radiation dominance will not be obtained.
In the present model, however, a radiation dominant era is recovered
as we will see.

Assuming the scalar field dominance, we find  the Friedmann equation
(\ref{fr1}) as
\beqn
H=\frac{1}{6m_5^{\;3}} \left[ \frac{1}{2}\dot{\phi}^2 +
\mu^{\alpha+4}\phi^{-\alpha} \right],
\label{frqsde}
\eeqn
while the equation for the scalar field (\ref{sde}) is
\beqn
\ddot{\phi}+3H\dot{\phi}-\alpha\mu^{\alpha + 4}\phi^{-\alpha-1}=0.
\label{ivsdesd}
\eeqn
Inserting Eq. (\ref{frqsde}) into Eq.
(\ref{ivsdesd}), we find a second order differential equation for $\phi$.
The asymptotic behavior of the solution can be
   classified into three cases:  (a) slow rolling (a
potential term dominant) solution $\left( \alpha < 2 \right)$,  (b)  a
solution in which the potential term balances with the  kinetic term $\left(
\alpha = 2
\right)$, and  (c) a kinetic term dominant solution $\left( \alpha > 2
\right)$.

First, assuming a slow rolling condition $(\dot{\phi}^2 \ll V$
and $|\ddot{\phi}| \ll H|\dot{\phi}|, |V'|)$,  Eqs.  (\ref{frqsde}) and
(\ref{ivsdesd}) are
\beqn
H&=&\frac{1}{6m_5^{\;3}} \mu^{\alpha+4}\phi^{-\alpha} ,
\label{frqsdesr}\\
3H\dot{\phi}&-&\alpha\mu^{\alpha + 4}\phi^{-\alpha-1}=0.
\label{ivsdesdsr}
\eeqn
  From Eqs. (\ref{frqsdesr}) and (\ref{ivsdesdsr}), we obtain  $\phi
\dot{\phi} =2\alpha m_5^{\;3}$, finding an analytic  solution
\beqn
\frac{\phi}{m_5}=2\left( \alpha m_5 \right) ^{1/2 }
\left(t-t_0\right)^{1/2},
\label{sevsr}
\eeqn
where $t_0$ is an integration constant.
With this solution, we have $\dot{\phi}^2  \propto t^{-1} $ and $
   V  \propto t^{-\alpha/2}$.
In order to guarantee the
slow-rolling conditions,  we have to require  $ \alpha < 2$.
   In this case,
the Universe expands as
\beqn
a = a_0 \exp\left\{ \left[H_0  \left(t-t_0 \right)\right]^{(2-\alpha
)/2}\right\},
\label{sqrtif}
\eeqn
where a constant $H_0$ is given by
\beq
\left({ H_0\over m_5}\right)^{(2-\alpha )/2}=
{1\over \left[3\left(2-\alpha
\right)
\left( 2\sqrt{\alpha}\right)^\alpha\right]}\left( {\mu\over
m_5}\right)^{\alpha+4}.
\eeq
Note that this solution describes an inflationary evolution, whose
expansion is weaker than the conventional exponential inflation, but
stronger than the power-law type.
Although this solution does not play into anything with a
quintessence scenario, it may be interesting to discuss a spectrum of
density perturbations for such an inflationary scenario. The results
will be published elsewhere.

Secondly, we assume a kinetic term dominance.
Eqs. (\ref{frqsde}) and (\ref{ivsdesd}) are now
\beqn
H&=&\frac{1}{12m_5^{\;3}}\dot{\phi}^2 ,
\label{frqsdekd}
\\
\ddot{\phi}&+&3H\dot{\phi}=0.
\label{ivsdesdkd}
\eeqn
Combining Eqs. (\ref{frqsdekd}) and (\ref{ivsdesdkd}), we have
$\ddot{\phi} =-(1/4m_5^{\;3})\dot{\phi}^3$, finding a solution as
\beqn
\frac{\phi}{m_5}=\pm 2\sqrt{2}m_5^{1/2}\left(t-t_0\right)^{1/2}
+\frac{\phi_0}{m_5},
\label{sevkd}
\eeqn
where $t_0$ and $\phi_0$ are integration constants.
For the solution with
$+$ sign,  the potential term decreases as $V  \propto
t^{-\alpha/2}$, while the kinetic  term drops as $\dot{\phi}^2
   \propto t^{-1}$. Hence, if $\alpha > 2$,  the kinetic
term  dominance condition is preserved.
  From Eq.  (\ref{frqsdekd}), we find that the Universe expands as
\beqn
a = a_0 (t/t_0)^{1/6},
\label{sckd}
\eeqn
and then
the
energy density of the scalar field decreases as
$\rho_{\phi} \propto t^{-1} \propto a^{-6}$,
while
$
\rho_r \propto a^{-4}
$.
Therefore, with this solution, the scalar field energy decreases
faster than that of radiation, and eventually radiation dominance will be
reached.

   For the solution (\ref{sevkd}) with
$-$ sign, the scalar field evolves into 0 as $t$ approaches a critical
value $t_{\rm cr}$ large, climbing the potential as
$V
\rightarrow
\infty$.
However, before reaching this critical point, the potential becomes
dominant, and then the assumption of a kinetic dominance  is no longer
valid in this limit. With the previous analysis for the potential
dominance, we expect that this solution will also reach to radiation
dominant stage.  This is confirmed by numerical calculations.
As a conclusion, the asymptotic behavior for the case with
$\alpha > 2$ is described by kinetic term dominance of the scalar field,
followed eventually by radiation dominance.

For the remaining case of $\alpha = 2$, we find an analytic
solution\cite{maeda}, which is a  power
law expansion of the Universe
\beqn
a = a_0 (t/t_0)^p,
\label{plif}
\eeqn
with
\beqn
p=\frac{1}{6}\left[1+\frac{1}{8}\left(\frac{\mu}{m_5}\right)^6\right].
\label{plindex}
\eeqn
and
\beqn
\phi=2\sqrt{2} m_5^{3/2}t^{1/2}.
\label{sevpl}
\eeqn
The scalar field energy density evolves as
\beqn
\rho_{\phi} \propto t^{-1} \propto a^{-1/p}.
\label{sfedqrple}
\eeqn

If $\mu > 40^{1/6}m_5 \approx 1.85 m_5$, $p>1$, that is, we find a
power-law  inflationary solution.
The power-law inflation is, however, nothing to do with quintessence,
although it may be interesting in the early universe.
If  $\mu < 4^{1/6}m_5 \approx 1.26 m_5$, $p < 1/4$, and then the scalar
field energy decreases faster than that of radiation.

\begin{table}[h]
\begin{tabular}{|@{}c || @{}c  || @{}c|@{}c|@{}l|}
    $\alpha$      & $\mu$         & ~initial~&~fate~&~~feature\\
\hline\hline
    \raisebox{-1.5ex}[0pt]{$\alpha < 2$}   &   \raisebox{-1.5ex}[0pt]{any
values}          & \raisebox{-0.5ex}[0pt]{S} &\raisebox{-1.5ex}[0pt]{S}&
\raisebox{-1.5ex}[0pt]{\footnotesize{~inflation}}
\\
\cline{3-3}
  &     &   \raisebox{-0.5ex}[0pt]{R}      & &
\\
\hline
  & $\mu/m_5\geq 40^{1/6}$ &
        &\raisebox{-1.5ex}[0pt]{S} &\footnotesize{~PL
inflation~}\\
\cline{2-2}
\cline{5-5}
  $\alpha = 2$ & $~4^{1/6}  <\mu/m_5 < 40^{1/6}~$ &   S   &
  &\footnotesize{~D.E}\\
\cline{2-2}
\cline{4-5}
   & $\mu/m_5 < 4^{1/6}  $ &   &
\raisebox{-1.5ex}[0pt]{R}&\raisebox{-1.5ex}[0pt]{\footnotesize{~scaling}}
\\
  \cline{2-3}
    & any values  & R        &
&\\
\hline
   ~\raisebox{-1.5ex}[0pt]{~$2 < \alpha < 6$}  &
\raisebox{-1.5ex}[0pt]{any values}   & \raisebox{-0.5ex}[0pt]{S}
&\raisebox{-1.5ex}[0pt]{R}
&\raisebox{-1.5ex}[0pt]{\footnotesize{~const}}\\
\cline{3-3}
  &    & \raisebox{-0.5ex}[0pt]{R} &  & \\
\hline
   \raisebox{-1.5ex}[0pt]{$\alpha\geq 6$} & \raisebox{-1.5ex}[0pt]{any
values}    & \raisebox{-0.5ex}[0pt]{S}  &\raisebox{-1.5ex}[0pt]{R}
&\raisebox{-1.5ex}[0pt]{\footnotesize{~kinetic}}\\
\cline{3-3}
  &
  &\raisebox{-0.5ex}[0pt]{R} &  &
\end{tabular}
\caption[table1]{\footnotesize{The fate of a scalar field for each
initial condition, where S and R denote
scalar field  dominance ($\rho_\phi \gg \rho_{\rm r}$)
and   radiation dominance ($\rho_{\rm r} \gg \rho_\phi $),
respectively. The ``scaling" means the scaling solution ($\Omega_\phi$
=constant). The asymptotic behavior of
$\rho_{\phi}^{(P)}/\rho_{\phi}^{(K)} =  {\rm
constant}$ is described by ``const", while    the
kinetic dominance of  $\rho_{\phi}^{(K)} \gg \rho_{\phi}^{(P)}$ is
denoted by ``kinetic". ``PL" is power-law. ``DE" is decelerating
power-law expansion.}}
\vskip .2cm
\end{table}
\noindent
Hence  radiation dominance will eventually be obtained.
For $4^{1/6}m_5<\mu < 40^{1/6}m_5 $, we can easily show that  this
solution is a global attractor and the ratio of  radiation energy to
that of the scalar field remains constant\cite{Future}.

We summarize the obtained analytic solutions and its fate
in Table 1.

   \subsection{Numerical analysis in $\rho^2$-dominant stage}

Now we study the dynamical property of the scalar field numerically and
show that the above analytic solutions are really attractors.
The  systematic analysis of the dynamical
properties of the scalar field in the quadratic dominant stage will be
given elsewhere\cite{Future}. Here we consider only the case of $2<\alpha
<6$ because we are interested in quintessence.
\begin{figure}
\epsfxsize = 3.3in
\epsffile{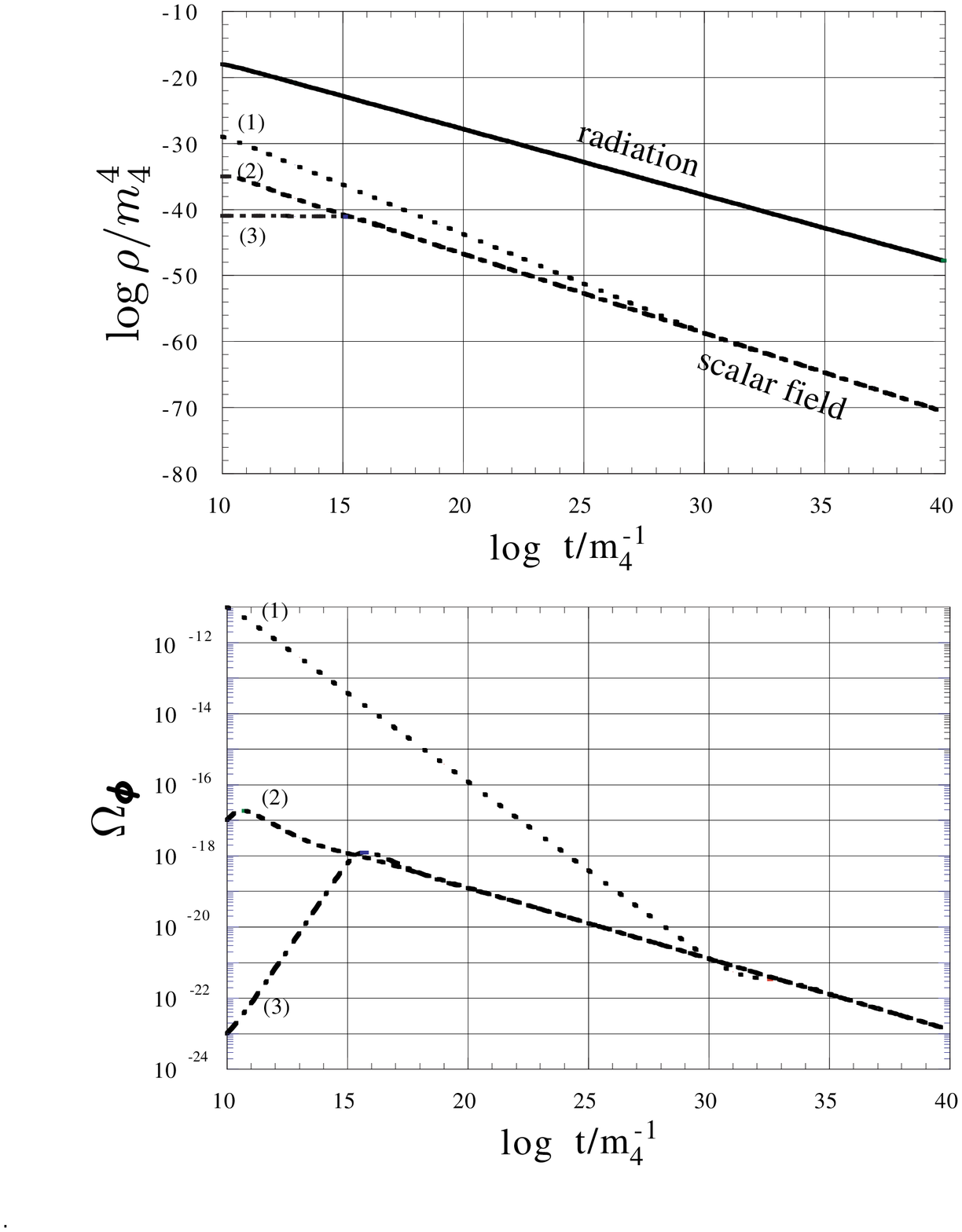}
\caption{
(Top) The time evolution of $\rho_{\rm r}$ and  $\rho_\phi$ starting from
a radiation dominant initial condition.
We set
$m_5=2.15 \times 10^{-3} m_4$, and
$\mu=1.0 \times 10^{-8} m_4$.
As for initial conditions, we set
$\rho_{\rm r}=1.0\times 10^{-18}
m_4^{\;4}$,  $\rho_\phi^{~(K)}=0$ and $
\rho_\phi^{~(P)}  = 1.0 \times 10^{-29} m_4^{\;4},
1.0 \times 10^{-35} m_4^{\;4},$ and $
1.0 \times 10^{-41} m_4^{\;4} $ for  (1), (2),
and (3), respectively.
If
$\rho_{\phi}$ is initially greater than the value of the attractor
solution (case  (1)), the kinetic term soon dominates and eventually
the attractor solution is reached. If
$\rho_{\phi}$ is initially less than the attractor's value (case
(3)),  the potential term dominates until the solution reaches the
attractor.
Thus, for  a wide range of initial
conditions ((1) $\sim$ (3)),
$\rho_\phi$ reaches  that of the  attractor solution.
(Bottom) Corresponding density parameter of the scalar field
($\Omega_\phi$)
   decreases after the attractor solution is reached.
}
\label{Fig1}
\end{figure}

\begin{figure}
\epsfxsize = 3.3in
\epsffile{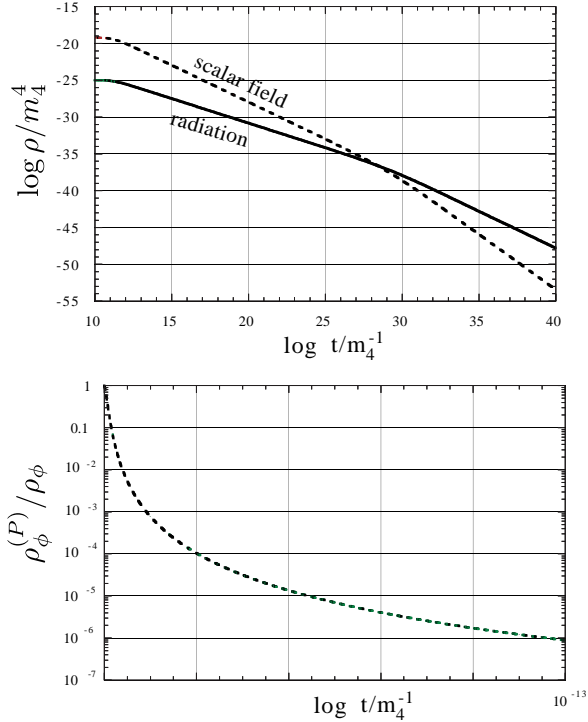}
\caption{
(Top) The time evolution of $\rho_{\rm r}$ and  $\rho_\phi$ 
from the scalar-field  dominant initial condition. The kinetic
term soon dominates the energy density of the scalar field, and
radiation  dominance is eventually reached.
As for initial conditions, we set
$\rho_{\rm r} =1.0\times 10^{-25} m_4^{\;4}$,
  $\rho_\phi^{~(K)}=9.35 \times 10^{-20} m_4^{\;4}$,
$\rho_\phi^{~(P)}=1.0 \times 10^{-17} m_4^{\;4},$
(Bottom) The time evolution of the ratio of the potential term of the
energy density of a scalar field to the total total energy density of the
scalar field. This suggests that the kinetic term dominates the potential
term soon.
We set
$m_5=2.15 \times 10^{-3} m_4$, and
$\mu=1.0 \times 10^{-8} m_4$.}
\label{Fig2}
\end{figure}

First, we show that  the solution (\ref{sevkpc}) is  a unique
attractor for plausible initial conditions. We depict  the evolution of
$\rho_{\rm r}$ and
$\rho_\phi$ for various initial  conditions in Figs.\ref{Fig1}
  and \ref{Fig2}   for $\alpha=3$.
  The figures in Fig.\ref{Fig1} show that
for a wide range of
initial conditions, the energy density of the scalar  field approaches
that of the attractor solution (\ref{sevkpc}).
We also find  that even starting from
scalar field dominance, the universe eventually evolves  into the
radiation dominant stage (see Fig. 2).
In fact, the kinetic term of a
scalar field
   always turns out to be dominant even if we start with a potential
dominance. Since
$\rho_{\phi} \propto a^{-6}$ in the case of the kinetic term dominance and
$\rho_{r} \propto a^{-4}$, radiation energy eventually  overcomes that of
the scalar field and the Universe evolves into a radiation
dominant era. As a result, the attractor solution is always reached for
any initial conditions.
    We   find similar results for any value of
$\alpha$ in
$2<\alpha <6$.
We may conclude that the  solution (\ref{sevkpc}) is a unique attractor in
the present dynamical system.

Once the  attractor is reached, the scalar field energy decreases
faster than that of  radiation, which is a most
interesting feature in the brane
quintessence scenario.
It is worthwhile noting that this  potential in the
conventional cosmology without the quadratic term, the  quintessence
scenario does not work if the scalar field energy initially dominates
that of
radiation\cite{tracker solutions}.
Next, we  shall  discuss a more natural quintessence scenario
in the brane world.

   \subsection{Quintessence scenario}

Now we are ready to  discuss a quintessence scenario
in the brane world\cite{maeda}. We assume   $2 < \alpha < 6$.
Using two attractor
solutions (one in the $\rho^2$-dominant stage and the other in the
conventional universe), we show a successful and natural scenario.
Since the quintessence solution in the conventional universe model is
an attractor, our solution should  also recover the same trajectory
after the quadratic term decreases to be very small. We have 
confirmed this numerically.
The main difference is that we can include not only radiation dominant
initial conditions  but also scalar-field dominant initial conditions
for a successful scenario.
In the numerical analysis,   to evaluate the
present value of the density parameter of a scalar field,
we include the
matter fluid as well as radiation and scalar field.

\vspace*{0.3cm}
$\left(1\right)$ a scenario
\vspace*{0.3cm}

First, we shall overview a quintessence scenario using attractor
solutions\cite{maeda}.  We introduce
$t_s$ (the cosmic time when the attractor solution in
$\rho^2$ dominant stage is  reached), $t_c$ (when the
$\rho^2$-term drops just below the conventional density term),
$t_{\rm NS}$(nucleosynthesis), $t_{\rm eq}$(when radiation energy
density becomes equal to matter density),
$t_{\rm dec}$(the decoupling time) and $t_0$ (the present time).
If we approximate the evolution of the Universe by the attractor solutions
in each stage, we find the analytic solution for the scalar field as
follows.
Normalizing the variables by 4-dimensional Planck mass scale $m_4$, we
find that the energy density
$\rho_\phi$ in each stage is described  very simply as
\beqn
\frac{\rho_\phi}{m_4^4}=
\left[\alpha(\alpha+2)^2\right]^{-{\alpha\over \alpha+2}}F(\alpha)
\left({\mu
\over m_4}\right)^{2(\alpha+4)\over (\alpha+2)} (m_4
t)^{-\frac{2\alpha}{\alpha+2}},
\label{nqui01}
\eeqn
where $F(\alpha)$ is a dimensionless constant defined only by $\alpha$.
In the $\rho^2$-dominant stage,
\beqn
F(\alpha)=(\alpha+2)\left({2\over 6-\alpha}\right)^{2\over \alpha+2}.
\label{quadra_attractor}
\eeqn
and in the conventional universe,
\beqn
F(\alpha)&=&(5\alpha+12)\left[2(\alpha+6)\right]^{-{2\over \alpha+2}}
\nonumber
\\ && ~~~~~~~~~~{\rm
(radiation ~dominant ~era)}
\label{radiation_attractor}
\\ &=&2^{3\alpha +2 \over
\alpha+2}(\alpha+2)(\alpha+4)^{-{2\over \alpha+2}} \nonumber \\
&&~~~~~{\rm (matter ~dominant
~era)}.
\label{matter_attractor}
\eeqn

Since the attractor solutions are independent, when the Universe shifts
from one attractor solution to the other one, we expect a discrepancy in
the energy density. However, since the difference in
$\rho_\phi$ in each stage  appears only in the factor $F(\alpha)$, the
discrepancy between two attractor solutions is given by $\alpha$.
We can easily check that the ratio of $F$ at the $\rho^2$-dominance to
that in the conventional radiation dominance is about 0.5 to 1 unless
$\alpha\approx 6$.  Note that the ratio of  radiation
dominant case to matter dominant case is about 0.8 for any values of
$\alpha$.
Hence,
when $t=t_c$, there is a  little discrepancy between the scalar field
energy  densities  estimated by two attractor  solutions.

Although the energy density of a scalar field changes quite similarly in any
stages, the radiation energy shows a big difference between
$\rho^2$-dominant stage and the conventional universe.
In fact, the radiation density decreases as $a^{-4}$, but the scale factor
changes as $a\propto t^{1/4}$ in the $\rho^2$-dominant stage in contrast
with  $a\propto t^{1/2}$ in the conventional radiation dominant era.
Therefore, when we discuss the density parameter of a scalar field
$\Omega_\phi \sim \rho_\phi/\rho_{\rm r}$, its behavior in the
$\rho^2$-dominant stage is completely different  from that in the
conventional universe.

\beqn
\Omega_\phi&\propto& t^{(2-\alpha)/(2+\alpha)} \sim
a^{(2-\alpha)/(2+\alpha)}~~~(\rho^2{\rm -dominant})\nonumber \\
&\propto&t^{4/(2+\alpha)} \sim a^{8/(2+\alpha)}~~~({\rm
radiation~dominant})\nonumber
\\ &\propto&t^{4/(2+\alpha)} \sim a^{6/(2+\alpha)}~~~({\rm
matter~dominant})
\eeqn
Since the radiation energy must be continuous, if we ignore the above
small discrepancies at $t_c$  and $t_{\rm eq}$, we can estimate the
density parameter $\Omega_\phi$ as
\beqn
\Omega_\phi &=& \Omega_\phi^{~(s)}\times
\left({a_c\over a_s}\right)^{-{\alpha-2\over 2\alpha+2}}\times
\left({a_{\rm eq}\over a_c}\right)^{8\over \alpha+2}\times
\left({a_0\over a_{\rm eq}}\right)^{6\over \alpha+2}\nonumber \\
&=&\Omega_\phi^{~(s)}\times
\left({T_s\over T_c}\right)^{-{\alpha-2\over 2\alpha+2}}\times
\left({T_c\over T_{\rm eq}}\right)^{8\over \alpha+2}\times
\left({T_{\rm eq}\over T_0}\right)^{6\over \alpha+2},
\eeqn
where $\Omega_\phi^{~(s)}$ is the density parameter when the attractor
solution is reached.
For a successful quintessence scenario, we require that the present
value of the density parameter of the scalar field is
$\Omega_\phi \sim 0.7$.

Before finding a constraint, it may be useful to  confirm the above analysis by
numerical study.
This is because with the above analytic attractor solutions, we  cannot
properly
treat the transition between
$\rho^2$-dominant stage and the conventional universe.
We show one numerical result for $\alpha =5$.
We set $\mu=6.0 \times 10^{-14} m_4$ ,
$m_5=6.0 \times 10^{-14} m_4$for a successful quintessence.

\begin{figure}
\epsfxsize = 3.3in
\epsffile{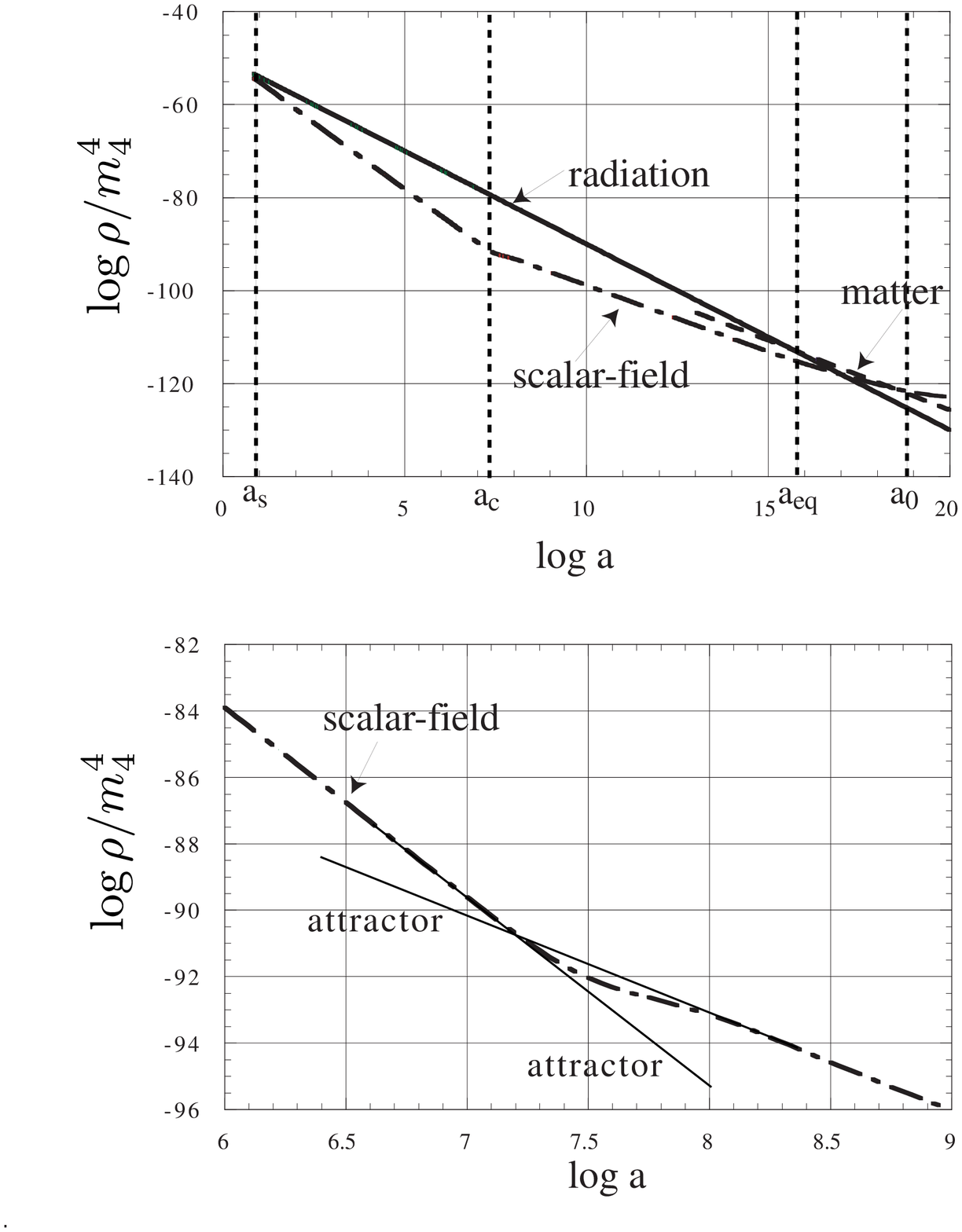}
\caption{(Top) The evolution of the energy densities in terms of a scale
factor.  We set $\mu=6.0 \times 10^{-14} m_4$, and
$m_5=6.0 \times 10^{-14} m_4$. As for the initial conditions, we set
$\rho_\phi^{~(K)}=3.82 \times 10^{-55} m_4^{\;4},
\rho_\phi^{~(P)}=2.50 \times 10^{-56} m_4^{\;4}, $ and
$\rho_{\rm r} =3.78\times 10^{-54} m_4^{\;4}$.
The matter fluid is also included to find the present matter dominant
universe as
$\rho_{\rm m} =5.43\times 10^{-69} m_4^{\;4}$.
The amount of matter fluid is
chosen in order to find the present universe.\\ (Bottom) The
enlargement of the top figure around the transition era from
$\rho^2$-dominant stage to the conventional universe. The thin solid lines
   are  attractor solutions both in $\rho^2$-dominant stage and in the
conventional universe.}
\label{Fig3}
\end{figure}

As initial conditions at $a_s=1$, we have chosen the attractor solution
Eq.(\ref{sfedqrrde}) in the
$\rho^2$-dominant stage, and solve the basic equations
(\ref{fr1}),(\ref{sde}), (\ref{ecr}) and (\ref{ecm}) including radiation
and matter fluids. In  Fig.\ref{Fig3}, we depict the
energy  densities of a  scalar field, radiation and matter  in
terms of a scale factor $a$. It turns out, as we expected, the discrepancy
at $a_c$
is very small and the evolution approximately 
follows the attractors in each stage
(attractor solutions as references).

\begin{figure}
\epsfxsize = 3.3in
\epsffile{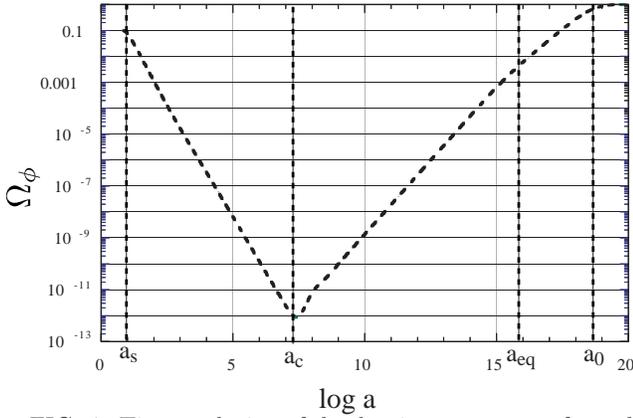}
\caption{Time evolution of the density parameter of a scalar field
for the same model in  Fig.\ref{Fig3}.}
\label{Fig4}
\end{figure}

 From Fig.\ref{Fig3},  we find that
$\rho_\phi\sim 10^{-90}m_4^{\;4}$ at $t_c$, $\sim
10^{-112}m_4^{\;4}$ at  decoupling time, and
$\sim 10^{-120}m_4^{\;4}$ at present (see Fig.\ref{Fig3}).
These small values guarantee a successful quintessence.
The conventional quintessence scenario (a tracking solution) is really
recovered when the quadratic energy density becomes small enough.
We also show the evolution of the density parameter of the
scalar field in Fig.\ref{Fig4}

\vspace*{0.3cm}
$\left(2\right)$ constraints to the extra-dimension
\vspace*{0.3cm}

We   discuss constraints for a natural and successful
quintessence.
We consider three constraints: nucleosynthesis, matter dominance at
decoupling time, and natural initial conditions, in order.

\vspace*{0.1cm}
$\left(a\right)$ nucleosynthesis
\vspace*{0.1cm}

One of the most successful results of the Big-Bang
standard cosmology is a natural  explanation of the present amount of
light elements. Therefore, in any cosmological models, a successful
nucleosynthesis provides a necessary constraint, which may be most
stringent. During nucleosynthesis, the universe  must expand as the
conventional radiation dominant era.
Therefore the transition from $\rho^2$-dominant stage to the conventional
universe must take place before nucleosynthesis.
This constraint gives a lower
bound for the value of
$m_5$. Introducing two temperatures, $T_c$ and $T_{\rm
NS}$, which correspond  to those at the transition time $t_c$  and at
nucleosynthesis time $t_{\rm NS}$, respectively, we describe  the
present constraint
   as  $T_c > T_{\rm NS} $, which implies
\beq
\rho_c>\rho_{\rm r}(t_c)={\pi^2\over 30}g_{\rm NS} T_{\rm NS}^{~4},
\label{NS_constraint}
\eeq
where $g_{\rm NS}$ is the degree of freedom of particles at nucleosynthesis.
Since  $\rho_c = 12m_5^6/m_4^2$ and $T_{\rm NS} \sim 1
{\rm MeV} $, the constraint (\ref{NS_constraint}) yields
\beqn
m_5 & > &1.6 \times 10^4(g_{\rm NS}/100)^{1/6} \nonumber\\
&\times& (T_{\rm NS}/1\:\:{\rm MeV})^{2/3} \:\:{\rm GeV}.
\label{qui10}
\eeqn
In the Planck unit, this constraint can be written as
$m_5> 10^{-14} m_4$.
If the  Randall-Sundrum II model is a fundamental theory,  in
order to recover the Newtonian  force  above 1mm scale in the brane
word, the 5-dimensional Planck mass is constrained as $m_5 \geq 10^8
\:\:{\rm GeV}\sim 4\times 10^{-11} m_4$\cite{RS}, which would be a
stronger constraint. However,
   the Randall-Sundrum II model could be  an  effective theory,
derived from   more fundamental higher-dimensional theories such as
Ho$\check{\rm r}$ava and E. Witten theory\cite{H-W}.
Thus, we adopt the above constraint here.

\vspace*{0.1cm}
$\left(b\right)$ The decoupling
\vspace*{0.1cm}

  From the observation  of the cosmic microwave background (CMB), we
have information  of the Universe at  $T\sim 4000$K, from which
we expect that inhomogeneity of the Universe  was about $10^{-5}$.  In
order to form some structure   from the  decoupling time to
the present,  the energy density of  matter fluid should be larger
than ``dark energy" (that of a scalar field) by a few orders of magnitude at
the decoupling time.

In $\rho^2$-dominant stage, the Friedmann
equation in the radiation dominant era is given by
\beqn
H=\frac{1}{6m_5^3}\rho_{\rm r},
\label{qui01}
\eeqn
and then  using $a\propto t^{1/4}$, we find
\beqn
\rho_{\rm r}=\frac{3}{2}  m_5^{~3} t^{-1}.
\label{qui02}
\eeqn
From, $\rho_c = 12 m_5^{\;6}/m_4^{\;2}$, we find  $t_c  =
m_4^{\;2}/(8m_5^{\;3})$.
In the conventional universe, $\rho_{\rm r} \propto t^{-2}$.
Then we have
\beqn
\rho_{\rm r} = \frac{3}{16} m_4^2 t^{-2},~~~{\rm for} ~~t_c<t<t_{\rm eq}
\label{qui12}
\eeqn

The energy density of matter fluid at decoupling time is given as
\beqn
\rho_{\rm m}(t_{\rm dec}) &= &\rho_{\rm m}(t_{\rm eq})  \left({a_{\rm
eq}\over a_{\rm dec}}\right)^3=\rho_{\rm r}(t_{\rm eq}) \left({T_{\rm
dec}\over T_{\rm eq}}\right)^3\nonumber \\
&=&\left({\pi^2\over 30}\right) g_{\rm eq}T_{\rm eq}^{~4}\times
\left({T_{\rm dec}\over T_{\rm eq}}\right)^3 .
\eeqn
As for the energy density of a scalar field,  assuming that  the attractor
(\ref{nqui01}) with (\ref{matter_attractor}) is reached and using Eq.
(\ref{qui12}),  we can estimate
$\rho_\phi(t_{\rm dec})$ in terms
of $\mu, T_{\rm eq}$ and $T_{\rm
dec}$ for each $\alpha$.

  $T_{\rm eq}$ is about $10^4{\rm K} < T_{\rm eq} < 10^5{\rm K}$.  In
order to  impose the most stringent constraint, we adopt  $T_{\rm eq} = 10^4$ K
here. Setting $T_{\rm dec} =4000$ K and  $T_{\rm eq} = 10^4$K, from the
constraint of $\rho_{\rm
m}(t_{\rm dec})> \rho_\phi(t_{\rm dec})$, we find  the upper bound for
the value of
$\mu$, i.e.
\beqn
\mu & <&  1.23 \times 10^{-16} m_4~~~{\rm for} ~~\alpha=3,,
\label{qui18}
\\
\mu & <&   1.31 \times 10^{-14}m_4~~~{\rm for} ~~\alpha=4,,
\label{qui19}
\\
\mu & <&   2.40 \times 10^{-13}m_4~~~{\rm for} ~~\alpha=5.
\label{qui20}
\eeqn

The value of $\mu$ is fixed if a scalar field dominates
now ($\Omega_{\phi}(t_0) \sim 0.7$).
However, since we do not know the value of $\mu$ from the viewpoint of
particle physics, we shall let its value be free.
We may discuss the naturalness of the present model, if we do not see
the coincidence problem.

\vspace*{0.1cm}
$\left(c\right)$ Initial condition
\vspace*{0.1cm}

About initial conditions, since a quintessence solution is an attractor,
we may not need to worry.
In fact, the conventional quintessence will be recovered even in  the
present model. What may be better in the present model is that a basin of
the attractor becomes larger. In particular, the conventional
quintessence will not work if a scalar field dominates initially, but it
will still work in the present model.

Nevertheless, here we will study about   natural initial conditions in the
present brane scenario.
Since we assume that a quintessence field $\phi$ is confined on the brane,
all energy scales including its potential should be smaller than the
5-dimensional Planck scale $m_5$;
\beqn
\mu \leq m_5.
\label{qui04}
\eeqn
The maximally possible energy density of radiation is also
about $ m_5^{\;4}$.

\begin{figure}
\epsfxsize = 3.3in
\epsffile{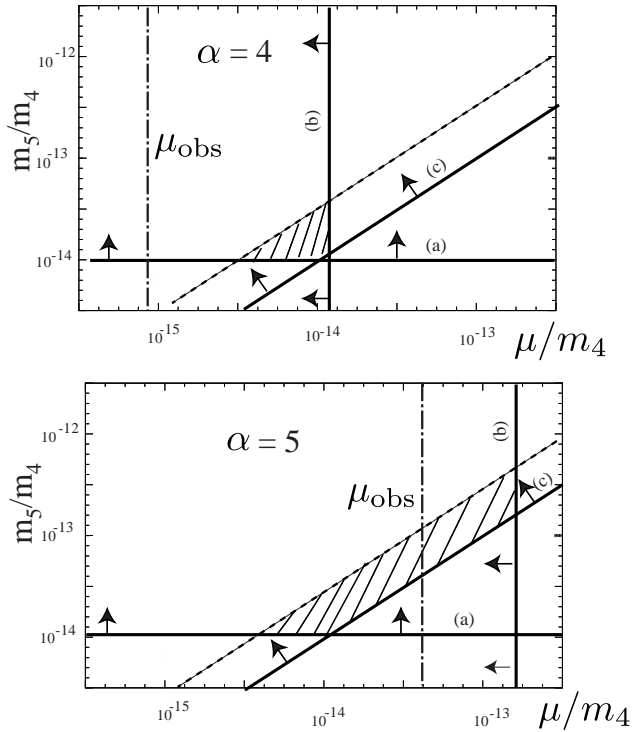}
\caption{Constraints in the $\mu$-$m_5$ parameter space. Three solid
lines are from (a) nucleosynthesis, (b) matter dominance at the
decoupling time, and (c) the energy scale smaller than the 5-D Planck
mass. The observation ($\Omega_\phi \sim 0.7$) fixes the value of
$\mu$ ($\mu_{\rm obs}$), which is given by the vertical dot-dashed lines.
If we assume that the potential energy of a scalar field is initially not
very small compared with its kinetic energy, a successful quintessence is
possible for a narrow range of parameters, which is shown by the   shaded
region.  }
\label{Fig5}
\end{figure}

With the constraints (a) and (b), we restrict two unknown parameters;
$m_5$ and $\mu$. In Fig. (\ref{Fig5}), we depict these constraints by
three solid lines in the
$m_5$ -$\mu$ parameter
space for $\alpha=4$ and 5.
 
Then if $\mu$ is fixed to find a scalar field dominance right now,
the 5-dimensional Planck mass $m_5$ is limited.
For example, for $\alpha=5$, if $\mu< 10^{-14}m_4$, the constraint for
$m_5$ is just coming from nucleosynthesis, while if $\mu > 10^{-14}m_4$,
$m_5>\mu$, which is  stronger than that of nucleosynthesis.

Here we invoke a further constraint which could be derived 
from natural initial
conditions.
What would be the natural initial conditions for a scalar field?
One plausible condition is
an equipartition of each energy density.
In this case, the radiation energy is larger than that of the scalar field
because the degree of freedom of all particles $g$ is larger than that of
the scalar field.
How about the ratio of the kinetic energy to  total one of 
the scalar field
? In the conventional quintessence scenario,
the potential energy should be initially much smaller than the kinetic one.
In the present model, it is not the case.  This is because
the attractor solution in the $\rho^2$-dominant stage reduces
the density parameter of the scalar field.
Therefore, we may impose natural initial conditions for a scalar field.
   To be more
concrete, we focus on the potential by fermion condensation. After our
3-dimensional brane world is created, a fermion pair is condensed by a
symmetry breaking mechanism and it behaves as a scalar filed with a
potential
$V(\phi)$. In this case, we expect that the potential term should play an
important role from the initial stage.

We consider the cases for $\alpha=3, 4,$ and 5.
The  initial conditions for a scalar field  are   classified
into the following three cases. First,
if
the kinetic and potential energy of a scalar field are the same order of
magnitude, the attractor solution is reached soon, and then we expect
$\Omega_\phi^{(s)} \sim 0(g_s^{~-1})$, where $g_s$ is a degree of freedom
of particles at
$a_s$.
Secondly, if the potential energy is dominant, it will not change so much
before reaching the attractor solution, and then we  expect that
$\Omega_\phi^{(s)}  \gsim 0(g_s^{~-1})$.
Thirdly, if the kinetic energy  is larger than the potential one, it will
decay soon, finding an attractor solution, then $\Omega_\phi^{(s)}  \lsim
0(g_s^{~-1})$, unless the kinetic energy dominates a lot, which we do not
assume here.
Therefore, a natural initial condition predicts  $\Omega_\phi^{(s)}  \sim
0(g_s^{~-1})$

From
Eqs. (\ref{nqui01}) and (\ref{qui02}), we find
\beqn
\Omega_\phi^{(s)}=G(\alpha)
\left({\mu
\over m_5}\right)^{2(\alpha+4)\over (\alpha+2)} (m_5
t_s)^{-\frac{2\alpha}{\alpha+2}},
\eeqn
where $G(\alpha)=2^{{\alpha+4\over
\alpha+2}}\alpha^{-{\alpha\over
\alpha+2}}(\alpha+2)^{-{\alpha-2\over
\alpha+2}}(6-\alpha)^{-{2\over
\alpha+2}}/3$.

Since $\rho_{\rm r}(t_s) \lsim m_5^4$, we have a constraint that
$m_5 t_s \gsim 1.5$ from Eq. (\ref{qui02}).
This constraint with $\Omega_\phi^{(s)}  \sim
0(g_s^{~-1})$ and $g_s \sim 10^3$ gives the lower bound for $\mu/m_5$ as
\beqn
\mu & \gsim & 0.131m_5
   ~~~{\rm for} ~~\alpha=3,
\label{qui06}
\\
\mu & \gsim &0.140m_5
   ~~~{\rm for}  ~~\alpha=4,
\label{qui07}
\\
\mu & \gsim &0.146m_5
~~~{\rm for}  ~~\alpha=5.
\label{qui08}
\eeqn

With the previous constraint $\mu<m_5$, we find a narrow strip in the
$\mu$-$m_5$ parameter space, which is shown by a shaded region in
Fig.(\ref{Fig5})
The allowed region gets  smaller for smaller values of
    $\alpha$. In particular, we find that  no region is allowed for
$\alpha\leq 3$.
Therefore,  the present model prefers a rather large value of
$\alpha$.

The values of $\mu_{\rm
obs}$,
which explains the observed value of the dark energy now, are
\beqn
\mu_{\rm obs}  &\sim& 6.25 \times 10^{-18}m_4 ~~~{\rm for}~~\alpha=3,
\label{qui21}
\\
\mu_{\rm obs}   &\sim& 8.75 \times 10^{-16}m_4~~~{\rm for}~~\alpha=4,
\label{qui22}
\\
\mu_{\rm obs}   &\sim& 4.06 \times 10^{-14}m_4~~~{\rm for}~~\alpha=5,
\label{qui23}
\eeqn
With these values, we find that there are no natural ranges for $\alpha
\leq 4$. For $\alpha\gsim  5$, the 5-dimensional Planck scale  is strictly
constrained from the observation because the  allowed region is very
narrow. For example, for
$\alpha=5$, we find
\beqn
4.06 \times 10^{-14}m_4 \lsim  m_5 \lsim  2.78 \times 10^{-13}m_4.
\eeqn
%

   \section{EXPONENTIAL POTENTIAL}
\label{sec.4}

Next we   investigate an exponential potential model, i.e.
\beq
V(\phi)=\mu^4\exp\left[-\lambda \frac{\phi}{m_4}\right],
\eeq
which is  another
typical potential for a quintessence
\cite{Ratra-Peebles},\cite{scaling  solutions}.
This type of potential is often found in unified theories of fundamental
interactions of particles such as supergravity
theory\cite{expotentilorigin}.

Within the conventional universe, this
potential shows an interesting property, although in itself it may not
provide a successful quintessence scenario.
We first recall a few  results\cite{Ratra-Peebles},\cite{scaling
solutions}.

Suppose that a spatially flat FRW universe  evolves with a scalar  field
$\phi$ and  a background fluid of an
equation of state $p_B=w_B \rho_B$.
There exist just two possible
attractor solutions, which show quite different late time properties,
depending on the values of $\lambda$ and $w_B$ as follows:

1. For $\lambda^2 > 3(w_B+1)$,  the
scalar field mimics a barotropic  fluid with $w_\phi \equiv
p_\phi/\rho_\phi = w_B$, and the relation
$\Omega_\phi \simeq \rho_\phi / \rho_B =3(w_B + 1)/\lambda^2$ holds,
where $\Omega_\phi$ is the density parameter of the
scalar field.

2. If $\lambda^2 < 3(w_B+1)$,  the late time attractor is a scalar
field dominant solution ($\Omega_\phi=1$)  with $w_\phi
=-1+\lambda^2/3$.

Case 1 is the so-called scaling solution. If it
is obtained in a radiation dominated era,  a successful nucleosynthesis
   is possible  for $\lambda^2 > 20$.  However, the present
observations of a scalar field dominance ($\Omega_\phi\simeq 0.7$) cannot
be explained by this type of solution.
Case 2 is preferred in the context of a quintessence scenario, but a
scalar field  behaves just as a cosmological constant in its evolution.
Then, in order to explain
$\Omega_\phi^{~(0)}\simeq 0.7$, an extreme fine-tuning in a choice of the
initial value of a scalar field or in a mass scale $\mu$ is required
just as the case of a cosmological constant.
Therefore, some modification for this type of potential has been done by
several authors for a successful quintessence\cite{Grad},
\cite{Ex-exponential}.

In the present paper, we study the effects of the quadratic term and
see whether a natural initial condition is found.
Hence, we will not analyze each modified potential quintessence model, but
rather study the universal properties which are found in an exponential
type potential. In particular, we are interested in case 2
above and see  whether a fine-tuning is loosened by the present scenario.

The organization of this section is as follows:
   first, as in the previous section, we focus on the $\rho^2$-dominant
stage, and  we  present an attractor solution which is expected to be
found as its asymptotic behavior.  This is confirmed by numerical
analysis.  Then, we discuss the possibility to improve a quintessence
scenario by the effects of the quadratic term.

   \subsection{Analytic solutions in the $\rho^2$-dominant stage }

Since $m_5$ rather than $m_4$ is a fundamental parameter in the present
model, it may be natural to introduce a  new parameter $\tilde{\lambda}
\equiv (m_5/m_4) \lambda$, and to write the potential in  the form of
\beq
V(\phi)=\mu^4\exp\left[-\tilde{\lambda}
\frac{\phi}{m_5}\right].
\eeq
In the brane world scenario, the value of $\tilde{\lambda}$ is expected
to be of order unity, unless the potential is coming from other physical
origin.

We discuss two  initial conditions, a radiation dominant initial
condition and a  scalar-field dominant one, in that order.

\vspace*{0.3cm}
$\left(1\right)$   radiation dominant initial condition
\vspace*{0.3cm}

In the case of a radiation dominant era, the scale factor expands as $ a
\propto t^{1/4}$ and the  equation for the scalar field  (\ref{sde})  is
now

\beqn
\ddot{\phi}+\frac{3}{4t}\dot{\phi}
-\left(\frac{\tilde{\lambda}\mu^4}{m_5}\right)
\exp\left[-\tilde{\lambda}\frac{\phi}{m_5}\right]=0.
\label{equi01}
\eeqn

Since the exponential potential drops much faster than the inverse-power
potential, unless $\phi\ll m_5/\tilde{\lambda}$, we expect  that the
kinetic energy dominant solution is  asymptotically found as the case of
the inverse-power  potential with $\alpha >6$. Hence,
assuming a kinetic energy dominant condition, we analyze the equations.
Ignoring the potential term in Eq.
(\ref{equi01}), we find
\beqn
\ddot{\phi}+\frac{3}{4t}\dot{\phi}=0,
\label{equi02}
\eeqn
which leads to the evolution of a scalar field as
\beqn
\phi \propto t^{1/4}.
\label{equi03}
\eeqn

  From this solution, $\rho_{\phi}^{~ (K)}
\propto t^{-3/2}$ and $\rho_{\phi}^{~ (P)} \propto \exp(-t^{1/4})$ ,
where
$\rho_{\phi}^{~ (K)} $
   and $\rho_{\phi}^{~ (P)}$ denote a kinetic  term of a scalar field  and
a potential one,
   respectively. This behavior confirms that the above kinetic-term
dominant solution gives an asymptotic behavior of the scalar field.

   However, for the case of $\phi\ll m_5/\tilde{\lambda}$, in
particular for an extremely small value of
$\tilde{\lambda}$, this potential behaves almost  the same as a
cosmological constant,
   and then the energy density of the scalar field will soon dominate
radiation, although we cannot give its critical value quantitatively.
However, as we will see later, it will again start to evolve into a
radiation dominant solution as long as the $\rho^2$-term dominates.
If the conventional universe is recovered before reaching the radiation
dominant stage, then the radation dominance will never be obtained because the
scalar field dominant solution is the attractor.

\vspace*{0.3cm}
$\left(2\right) \phi$-dominant initial condition
\vspace*{0.3cm}

If the scalar field dominates initially,  the Friedmann equation
(\ref{fr1}) is
\beqn
H=\frac{1}{6m_5^{\;3}} \left( \frac{1}{2}\dot{\phi}^2 +
\mu^4\exp\left[-\tilde{\lambda} \frac{\phi}{m_5}\right]\right),
\label{equi04}
\eeqn
while the equation for the scalar field (\ref{sde}) is
\beqn
\ddot{\phi}+3H\dot{\phi}-\left(\frac{\tilde{\lambda}\mu^4}{m_5}\right)
\exp\left[-\tilde{\lambda}\frac{\phi}{m_5}\right]=0.
\label{equi05}
\eeqn

We again expect  that the kinetic  term
dominant solution gives an   asymptotic behavior.
Assuming the kinetic term dominant condition, from Eqs.
(\ref{equi04}) and (\ref{equi05}), we find
\beqn
\ddot{\phi}+\frac{1}{4m_5^{\;3}}\dot{\phi}^3 =0,
\label{equi07}
\eeqn
which is the same as Eqs. (\ref{frqsdekd})  and (\ref{ivsdesdkd}),
because the potential term does play no role.
We find the same solution for a scalar field as before (Eq.(\ref{sevkd})),
and then the same result, i.e. this  gives an asymptotic solution.
Since  radiation energy decreases slower than
that of  a massless scalar field, the universe will eventually evolve
into a radiation dominant era, just as  discussed in the previous section.

However, as discussed before, if $\tilde{\lambda}$ is extremely small, the
potential does not decay so fast and then after recovering the
conventional universe, a scalar field will lead to inflation.

   \subsection{ Numerical analysis }

\begin{figure}
\epsfxsize = 3.3in
\epsffile{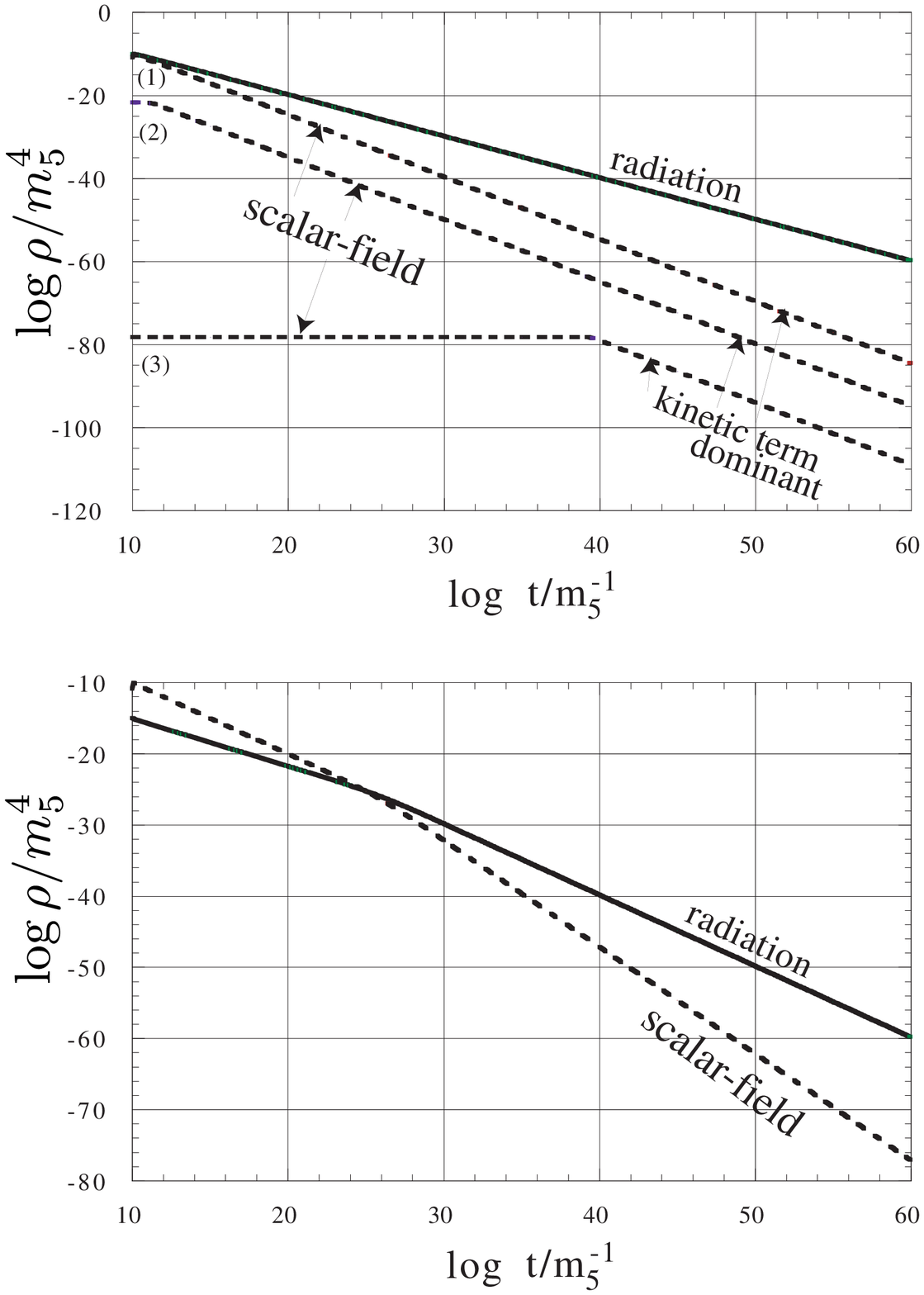}
\caption{ (Top) Time evolution of each energy density for
$\tilde{\lambda}=1.0$ in the  radiation dominant era. The kinetic term
dominant solution is realized for any initial conditions including the
case of potential  dominance. As for the initial conditions, we set
$\rho_{\rm r} =1.0\times 10^{-10} m_5^{\;4},
  \rho_\phi^{~(K)}=0.0,
\rho_\phi^{~(P)}=1.39\times 10^{-11} m_5^{\;4}, $ for (1),
$ \rho_\phi^{~(K)}=0.0,
\rho_\phi^{~(P)}=4.22 \times 10^{-22} m_5^{\;4}, $ for (2),
$ \rho_\phi^{~(K)}=0.0,
\rho_\phi^{~(P)}=6.71 \times 10^{-79} m_4^{\;4}, $ for (3).
(Bottom) Time evolution of each energy density
for $\tilde{\lambda}=1.0$, starting from the scalar-field dominant
initial condition.
As for the initial condition, we set
$\rho_{\rm r} =1.0\times 10^{-15} m_5^{\;4},
  \rho_\phi^{~(K)}=0.0,
\rho_\phi^{~(P)}=1.39\times 10^{-11} m_5^{\;4}.$
The kinetic term dominates the energy density of a scalar field
soon, and  the radiation dominance is eventually  reached.}
\label{Fig6}
\end{figure}

In order to show that the above kinetic term dominant solution in a
radiation dominant era is a unique attractor for any initial condition,
we present  numerical results.
In Fig.\ref{Fig6}, we depict  the evolution of each energy  density in
the quadratic term dominant stage.
In the top figure, we show the results for the initial condition of
radiation dominance.
We find that even for the  initial
conditions such that the potential term of a scalar field dominates the
kinetic one, we eventually find a kinetic dominant solution. In the
bottom, we also show the case
  of a scalar-field dominant initial condition.  The universe eventually
evolves into a radiation dominant era. Therefore, as long as
$\tilde{\lambda}$ is of order unity, the solution obtained above is found
asymptotically  both from the radiation dominant initial condition
and from  the scalar-field dominant initial condition.
We may conclude that radation dominance is a natural condition for
the $\rho^2$-dominant stage.

   \subsection{A quintessence scenario}

Although a pure exponential potential may not give a natural quintessence
model, we shall study whether or not a similar mechanism for the value
of
$\Omega_\phi$ in the $\rho^2$-dominant stage works.
If it works, it may provide a natural initial condition for a quintessence
model based on an exponential potential.
We discuss two cases; (a) $\tilde{\lambda} \sim O(1)$ and (b) $\lambda
\sim O(1)$ in order.

\underline{Case (a)} : Suppose  $\tilde{\lambda} \sim O(1)$ because the
potential causes  the 5-dimensional origin.
In this case, as we discussed above, the radiation dominant universe is
an attractor and the kinetic term dominates the potential for a scalar
field.
When the universe evolves into the conventional expansion stage,
the scalar field approaches an attractor of a scaling solution soon,
because $\lambda =(m_4/m_5)\tilde{\lambda} \gg 1$.
In fact, we find $\lambda > 10 ^{14}$ from the constraint on $m_5$ by
nucleosynthesis. The ratio of the scalar field energy to radation energy,
which  is fixed by
$\lambda$ as $\Omega_\phi \simeq \rho_\phi / \rho_B =3(w_B +
1)/\lambda^2$, turns out to be very small in the present model. Therefore,
this does not provide any quintessence model. In order to remedy it, we
may need an additional potential for quintessence models with exponential
type potentials in the conventional universe\cite{Ex-exponential}. For
example, suppose that the potential is $V(\phi) =\mu^4
\exp[-\tilde{\lambda}
\phi/m_5] +\mu_2^4 \exp[\lambda_2
\phi/m_4]$. If $\mu_2 \sim 10^{-30} m_4$, we find that min($V$) $\sim
\mu_2^4 \sim \rho_{\rm cr}$, which gives the present small cosmological
constant. However, an introduction of such an additional potential may
break naturalness in the present model.

\begin{figure}
\epsfxsize = 3.3in
\epsffile{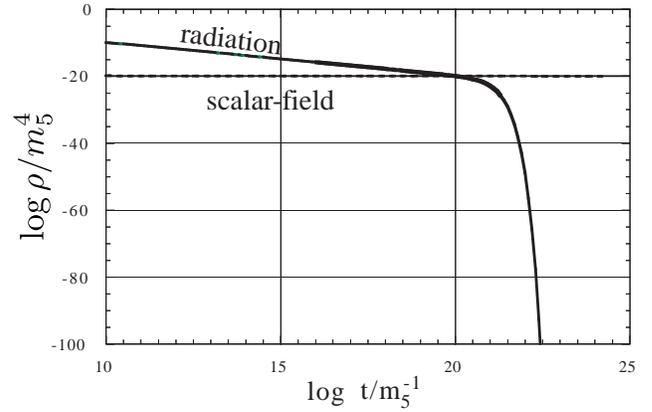}
\caption{ The time evolution of the energy densities for extremely small
$\tilde{\lambda}$, in which case  the scalar field behaves as the cosmological constant.
Even though starting from a radiation dominant initial condition, the scalar
field dominates the radiation, and inflation happens. In this case, we
take
$\tilde{\lambda}= 1.0 \times 10^{-14}$. As for the initial condition,  we
set
$\rho_{\rm r} =1.0\times 10^{-10} m_5^{\;4},
  \rho_\phi^{~(K)}=0.0,
\rho_\phi^{~(P)}=1.50\times 10^{-20} m_5^{\;4}.$
}
\label{Fig7}
\end{figure}

\underline{Case (b)} : If we have  $\lambda \sim O(1)$,i.e.
  $\tilde{\lambda} \sim
\frac{m_5}{m_4}\ll 1$,
the potential may behave as a cosmological constant in
$\rho^2$-dominant stage.
We show for the case with
  initial conditions that  $\rho_r
> \rho_ \phi $ (Fig.\ref{Fig7}). The energy density of the scalar field
remains constant and eventually dominates the  radiation, leading to the
inflationary stage, unless the conventional universe is recovered.
Hence this does not change the conventional quintessence model with
exponential potential. If  the  kinetic term
  dominates in the energy density of a scalar field, however,  the energy
density will decrease in time.

We then study this case in what follows.
We have to know the value of $V(\phi(a_c))$, which is approximately
the present value of a cosmological constant, because
the potential term remains almost constant as a cosmological constant
after the universe enters the conventional expansion stage.
The evolution for a scalar field in the $\rho^2$-dominant stage
is given by Eq.(\ref{equi03}), i.e. $\phi \propto a \propto 1/T$.
\beqn
V(\phi(a_c)) &\sim& V(\phi(a_s)) \exp \left[
-\lambda   \phi_c/m_4\right] \nonumber \\
&\sim& V(\phi(a_s))\exp\left[
-\lambda {\phi_s T_s\over  m_5^2} \left({m_5
\over m_4}\right)^{1/2}\right] .
\label{exp_pot}
\eeqn
  In order to find $V(\phi(a_c)) \sim \rho_{\rm cr}$, the exponent in the
r.h.s. in Eq.(\ref{exp_pot}) should be very large.
However, since $m_5\ll m_4$, the initial value of $\phi_s$ should be large
as
\beq
\phi_s \gsim 10^2  \left({m_4
\over m_5}\right)^{1/2} m_5 .
\eeq
This initial condition may not be natural.

We may conclude that for exponential type potential, a brane world
does not improve the quintessence scenario.

\section{KINETICALLY DRIVEN QUINTESSENCE}
\label{sec.5}

There is another type of quintessence model in the conventional
universe, in which the quintessential dynamics is driven solely by
a (non-canonical) kinetic term rather than by a potential
term\cite{K-Quintessence},\cite{K-essence} . It is called
"k-essence". Here we shall study it in the context of a brane world.

The model Lagrangian of a scalar field is given by
\beqn
S=\int d^4 x \sqrt{-g}\left[- K(\phi) X +L(\phi)
X^2\right],
\label{S_kinetic}
\eeqn
where $X={1\over 2} \left(\nabla
\phi\right)^2$.
If we introduce a new scalar field by
\beqn
\Phi =\int \sqrt{L\over |K|} d\phi ,
\eeqn
the action (\ref{S_kinetic}) is rewritten as
\beqn
S=\int d^4 x \sqrt{-g} f(\Phi)\left[-X +X^2\right].
\label{S_kinetic2}
\eeqn

Among them, the model defined by
\beqn
f(\Phi) = \mu^{4-\alpha}\Phi^{-\alpha} ,
\label{f_power}
\eeqn
provides a "tracking" solution\cite{K-Quintessence}.
$\mu$ is a typical mass scale of the system and will fix when
the scalar field dominates. We then  expect a similar or more
interesting feature in the
$\rho^2$-dominant stage. It may provide more natural initial
conditions as the inverse power-law potential discussed in
\S 3. Note that a scalar field
$\Phi$ has not  mass dimension but inverse-mass dimension.

Contrary to our
expectation, however, we do not find a solution for which the density
parameter of a scalar field decreases.
Instead, as
an attractor, we have a ``tracking solution" in which the density parameter
increases slower than that in the conventional universe.  Furthermore, we
find a "scaling" solution as a transient attractor in the radiation
dominant  era.

   \subsection{Analytic solutions}

\vspace*{0.3cm}
$\left(1\right)$ Model
\vspace*{0.3cm}

First, we shortly explain our quintessence model with Eqs.
(\ref{S_kinetic2}) and (\ref{f_power}). Assuming the FRW universe model,
the ``pressure",
$p_{\Phi}$, and the ``energy density", $\rho_{\Phi}$, of a quintessence
scalar field $\Phi$ is given by

\beqn
p_{\Phi}= \mu^{4-\alpha} \Phi^{-\alpha}\left(-X + X^2 \right),
\label{kqui01}
\eeqn
\beqn
\rho_{\Phi}= \mu^{4-\alpha} \Phi^{-\alpha}\left(-X + 3X^2 \right),
\label{kqui02}
\eeqn
where $X = (1/2)\dot{\Phi}^2$. Note that in order to guarantee the
positive energy
density of the scalar field , $\rho_{\Phi} $, it is necessary that
$\dot{\Phi}^2  >(2/3)$.

The field equation corresponding to (\ref{sde}) is
\beqn
\ddot{\Phi}(1-3\dot{\Phi}^2)+3H(1-\dot{\Phi}^2)\dot{\Phi}-
\frac{\alpha}{4\Phi}(2-3\dot{\Phi}^2)\dot{\Phi}^2=0,
\label{kqui03}
\eeqn
Since this equation has a reflection symmetry ($\Phi \leftrightarrow
-\Phi$), we discuss only the case of
$\Phi > 0$.

\vspace*{0.3cm}
$\left(2\right)$ Analytic solutions in the conventional cosmology
\vspace*{0.3cm}

First we  show  two  analytic solutions  in the
conventional cosmology. We assume the radiation dominant
era, even though a similar solution is found  in the
matter dominant era.

Then the scale factor expands as $a \propto t^{1/2}$ and the
equation for the scalar field is Eq.  (\ref{kqui03}) with $H=1/2t$.

We have two analytic solutions: (a) a tracking solution, which is exact
and is an attractor, and (b) a scaling solution, which is approximate and a
transient attractor.

\underline{ (a) a tracking solution}:
For Eq.(\ref{kqui03}) with $H=1/2t$ , we have
  an exact solution for $\alpha
< 2$ or for $\alpha>3$ as
\beqn
\Phi=\sqrt{\frac{2(3-\alpha)}{3(2-\alpha)}}t,
\label{kqui05}
\eeqn
which yields its energy density  as
\beqn
\rho_{\Phi}=\frac{\mu^{4-\alpha}}{2(2-\alpha)}
\left[\frac{2(3-\alpha)}{3(2-\alpha)}
\right]^{(2-\alpha)/2}
t^{-\alpha}\propto a^{-2\alpha}.
\label{kqui06}
\eeqn
For $\alpha < 2$, the energy density of the scalar  field decreases slower
than
that of radiation, i.e. the density parameter
$\Omega_{\Phi}$ increases as $a^{2(2-\alpha)}$. This
solution is the tracking solution.

For $\alpha > 3$, the energy density becomes negative, although the
density parameter $\Omega_{\Phi} = |\rho_\Phi/\rho_{\rm r}|$  decreases.
This  solution may not be interesting for a
quintessence scenario because the scalar field energy never dominates.

\underline{(b) a scaling solution}: Another interesting  solution is found
in the limit of
$\dot{\Phi}^2 \gg 1$. In this case, the equation for a scalar field
(\ref{kqui03}) with $H=1/2t$ is
\beqn
\ddot{\Phi}+\frac{1}{2t}\dot{\Phi}-
\frac{\alpha}{4\Phi}\dot{\Phi}^2=0.
\label{kqui07}
\eeqn
It is easy to find a solution for
Eq. (\ref{kqui07}),which is
\beqn
\Phi \propto t^{2/(4-\alpha)} ~~~{\rm and}~~~
\dot{\Phi} \propto t^{-(2-\alpha)/(4-\alpha)}.
\label{kqui08}
\eeqn
The coefficient is an integration constant and depends on the initial
condition. The energy density is
now
\beqn
\rho_{\Phi} \propto t^{-2} \propto a^{-4},
\label{kqui09}
\eeqn
  This   is nothing but a
scaling solution in a radiation dominant era.

We easily show that this solution is an attractor in the present
system with the approximation of $\dot{\Phi}^2\gg 1$.
However, this solution (\ref{kqui08}) shows that the approximation will be
eventually
broken because $\dot{\Phi}$ decreases. Hence, after this approximation
becomes invalid, the universe evolves into the tracking
solution.
Note that  for $\alpha <3$, $\dot{\Phi}$ increases and the approximation
is always valid, although the energy density is negative.

\vspace*{0.3cm}
$\left(3\right)$ Analytic solutions in the $\rho^2$-dominant stage
\vspace*{0.3cm}

In the $\rho^2$-dominant stage, assuming  the radiation dominant
era, i.e. the evolution of the  scale factor is $a \propto t^{1/4}$,
the equation for a scalar field is given by Eq. (\ref{kqui03}) with
$H=1/4t$.

The same as in the conventional universe, 
we have two analytic solutions:
(a) a tracking solution, which is exact and an attractor, and (b) a scaling
solution, which is approximate and a transient attractor.

\underline{ (a) a tracking solution}:
There is an exact solution  for
$\alpha < 1$ or for $\alpha>3/2$, which is
\beqn
\Phi=\sqrt{\frac{3-2\alpha}{3(1-\alpha)}} t,
\label{kqui11}
\eeqn
which yields the energy density of a scalar field as
\beqn
\rho_{\Phi}={\mu^{4-\alpha}\over
4(1-\alpha)} \left[\frac{3-2\alpha}{3(1-\alpha)}
\right]^{(2-\alpha)/2}
t^{-\alpha} \propto a^{-4\alpha}.
\label{kqui12}
\eeqn
The density parameter $\Omega_{\Phi} $ increases as
$a^{4(1-\alpha)}$ for
$\alpha < 1$. This solution is again a tracking solution, although the
rate of increase is smaller than that in the conventional universe.
There is no solution in which  $\Omega_{\Phi} $ decreases.
This is a big difference between the present model and
the model with
inverse-power potential. For the
case of
$\alpha > 3/2$,
$\rho_{\Phi}$  becomes negative.
Since the positivity of the energy density of a scalar field is not
guaranteed, it might be allowed if the Friedmann equation is not contradicted
as in the case of a radiation dominant era.
Although it might be interesting because the density parameter
$\Omega_{\Phi} $ decreases, it turns out that such a solution cannot
evolve into a tracking solution in the conventional universe
\cite{K-Quintessence}.
Hence, we do not discuss this solution in what follows.

\underline{(b) a scaling solution}: We have another solution
in the limiting case of
$\dot{\Phi}^2 \gg 1$.
  The equation for the scalar field
\beqn
\ddot{\Phi}+\frac{1}{4t}\dot{\Phi}-
\frac{\alpha}{4\Phi}\dot{\Phi}^2=0.
\label{kqui13}
\eeqn
has an analytic solution
\beqn
\Phi \propto t^{3/(4-\alpha)}~~~{\rm and }~~~\dot{\Phi} \propto
t^{-(1-\alpha)/(4-\alpha)},
\label{kqui14}
\eeqn
and
\beqn
\rho_{\Phi} \propto t^{-1} \propto a^{-4},
\label{kqui15}
\eeqn
which is a scaling solution.
Since $\dot{\Phi}$ decreases as Eq.(\ref{kqui14}), the approximation of
$\dot{\Phi}^2 \gg 1$ will be eventually broken.
This scaling solution is a transient attractor.
Then, a tracking solution will be finally reached for $\alpha <1$.
Since there is no attractor solution for $\alpha \geq 1$, the evolution of
a scalar field may depend on the initial conditions.
Once the universe evolves into the conventional expansion stage, however,
the scalar field begins to approach an attractor solution.

Next, we investigate the case of the scalar-field  dominant era.
The  Friedmann equation (\ref{fr1}) is given by
\beqn
H=\frac{1}{6m_5^{\;3}}\mu^{4-\alpha}\Phi^{-\alpha}\left(-X + 3X^2 \right).
\label{kqui16}
\eeqn

In the limit of  $\dot{\Phi}^2 \gg 1$, we find a power law solution as
\beqn
a &\propto &t^{1\over 4}
\label{kqui19} \nonumber \\
\Phi &= &{\Phi_0 \over \mu}(\mu t)^{3\over (4-\alpha)},
\eeqn
where a dimensionless constant  $\Phi_0$ is given by
\beqn
\Phi_0 = 2^{1\over 4-\alpha} \left(\frac{\mu}{m_5}\right)^{-{3\over
4-\alpha}}
\left(\frac{4-\alpha}{3}\right)^{4\over 4-\alpha}.
\label{kqui21}
\eeqn
Then the  energy density of the scalar field  is given by
\beqn
\rho_{\Phi} = \frac{3}{2} m_5^3 t^{-1} \propto a^{-4},
\label{kqui22}
\eeqn
which drops  as  the radiation energy.
Hence, once this solution is reached, contrary to the case of inverse-power
law potential,  the radiation never dominates the  scalar field.
Therefore, a scalar field dominant initial condition does not provide
a successful quintessence scenario.

   \subsection{ Numerical analysis}

\begin{figure}
\epsfxsize = 3.3in
\epsffile{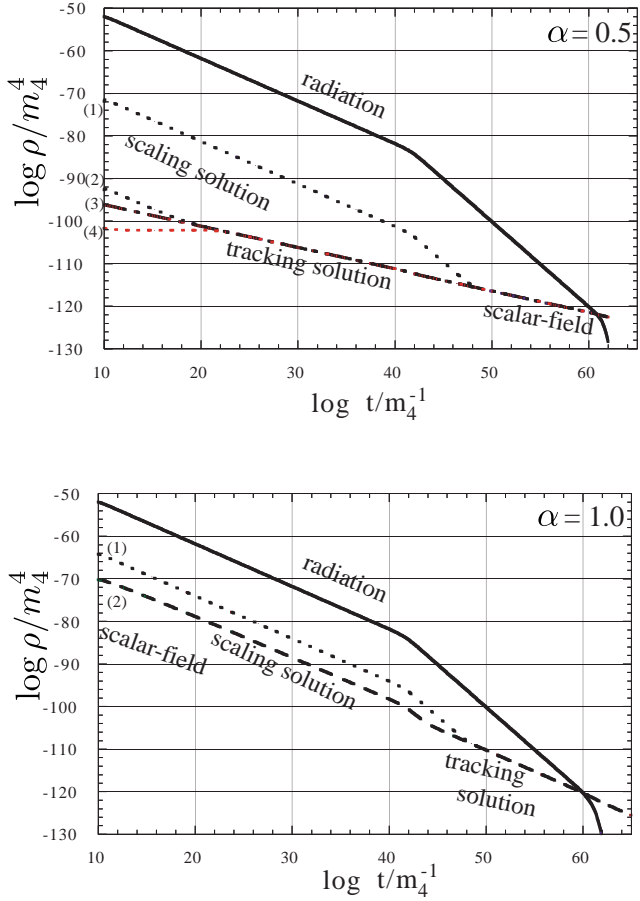}
\caption{(Top) Time evolution of each energy density for $\alpha
=0.5 $. We set $\mu = 1.0 \times 10^{-26} m_4, m_5= 1.0 \times
10^{-14}m_4$. As for initial conditions, we set
$\rho_{\rm r} =1.0\times 10^{-52} m_4^{\;4}$, and
$\rho_{\Phi} =1.24\times 10^{-72} m_4^{\;4}$ 
($\Phi=1.15\times 10^{10}
m_4^{\;-1}, X=6.67 \times 10^{11}$), 
$1.96\times 10^{-93} m_4^{\;4}$ 
($\Phi=1.15\times 10^{3} m_4^{\;-1},
X=6.67 \times 10^{-1}$), 
$6.21\times 10^{-97} m_4^{\;4}$
($\Phi=1.15\times 10^{10} m_4^{\;-1}, X=6.67 \times 10^{-1}$)
, and $1.95\times 10^{-102} m_4^{\;4}$
($\Phi=1.15\times 10^{21} m_4^{\;-1},
X=6.67 \times 10^{-1}$)
for (1), (2), (3) and (4), respectively.
It is easy to see that the tracking solution  is an
attractor, while the scaling solution  is a transient attractor.
(Bottom) The same figure as the top for $\alpha = 1$.
We set $\mu = 1.0 \times 10^{-20} m_4, m_5= 1.0 \times 10^{-14}m_4$.
As for initial conditions, we set
$\rho_{\rm r} =1.0\times 10^{-52} m_4^{\;4}$, and
$\rho_{\Phi} =5.77\times 10^{-65} m_4^{\;4}$
($\Phi=1.15\times 10^{4} m_4^{\;-1},
X=6.67 \times 10^{-1}$),
$5.77\times 10^{-71} m_4^{\;4}$
($\Phi=1.15\times 10^{10} m_4^{\;-1},
X=6.67 \times 10^{-1}$)
for (1), (2) , respectively.
There is no attractor solution in the $\rho^2$-dominant stage.
After the  conventional cosmology is recovered, the
scaling solution evolves into a tracking solution.}
\label{Fig8}
\end{figure}

In order to confirm the above  analysis,  we shall show  our numerical
results. The initial condition is
chosen such that the inequality
$\dot{\Phi}^2  > 2/3$ is satisfied, which guarantees   positivity of the
energy density of a scalar field. First, we show the case with
radiation dominant initial conditions in Fig.\ref{Fig8}. We find that a
tracking solution (\ref{kqui11}) or a scaling solutions (\ref{kqui14}) is
really an attractor or a transient one.

\begin{figure}
\epsfxsize = 3.3in
\epsffile{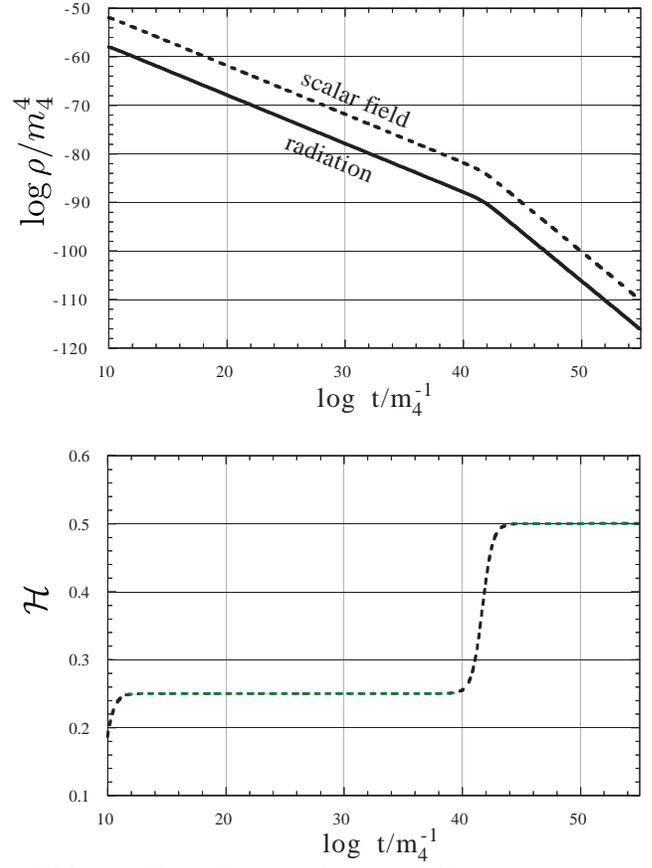}
\caption{(Top) Time evolution of the energy densities with
the scalar field dominant initial conditions for $\alpha=1$. 
As for
initial  conditions, we set
$\rho_{\rm r} =1.0\times 10^{-58} m_4^{\;4}$ and
$\rho_{\Phi}=1.12\times 10^{-52} m_4^{\;4},$
($\Phi=1.66\times 10^{19} m_4^{\;-1},
X=2.45 \times 10^{13}$).
Note that both energy
densities  drop as $\propto t^{-1}$
in the $\rho^2$-dominant era and $\propto t^{-2}$ 
in the conventional universe.
(Bottom) The change of the power of scale factor, ${\cal H}= Ht$. Even
in  a scalar field dominant universe, the cosmic expansion  is the same
as that of the radiation, i.e.  $a \propto t^{1/4}$ the  $\rho^2$-dominant
era and  $t^{1/2}$ in the conventional universe.
We set
$\mu = 1.0 \times 10^{-20} m_4, m_5= 1.0 \times 10^{-14}m_4.$}
\label{Fig9}
\end{figure}

Next,  we show  numerical results for  the scalar field dominant initial
conditions in Fig.\ref{Fig9}. The
energy density of the radiation never  dominates that of the scalar field,
and there exists  no radiation dominant era, which is excluded from the
constraint of   nucleosynthesis.

   \subsection{Constraint to the model}
Now, we discuss the value of the parameter $\mu$.
For a successful quintessence, i.e. in order for a scalar field to
  dominate the  energy density right now, we have to tune the value of
$\mu$. Naive  estimation gives
\beqn
\rho_{\Phi 0} = \mu^{4-\alpha}\Phi^{-\alpha}_ 0 \left(-X_0 + 3X_0^2 \right)
\simeq
\rho_{\rm (cr) 0},
\label{kqui23}
\eeqn
where $\Phi_0$ and $X_0$ are  the present values of $\Phi$ and $X$, and
$\rho_{\rm (cr) 0}$ is the present value of the critical density.

The present value of the scalar field is estimated by
a tracking
solution  in the matter dominant era, which  is
\beqn
\Phi=\sqrt{\frac{2(4-\alpha)}{8-3\alpha}}t.
\label{kqui24}
\eeqn
Since $H_0 \sim 2/3t_0$, we find
\beqn
{\Phi_0}^{-1}&=&\frac{3}{2} H_0 \sqrt{\frac{8-3\alpha}{8-2\alpha}}
\simeq H_0,\nonumber \\
X_0 &=&\frac{(4-\alpha)}{8-3\alpha} \sim O(1)
\label{kqui25}
\eeqn
for $0 < \alpha < 8/3$.
These equations with Eq. (\ref{kqui23}) fix $\mu$ and the
present value of the scalar field as
\beqn
\mu \sim 10^{(43\alpha-48)/(4-\alpha)} [{\rm GeV}],
\label{kqui26}
\eeqn
\beqn
\Phi_{0} \sim10^{43}[{\rm GeV}^{-1}].
\label{kqui 27}
\eeqn
If the mass scale $\mu$ is the same order of magnitude as the
five-dimensional Planck mass $m_5$, we have a constraint on the value
of $\alpha$ from nucleosynthesis, i.e. $m_5
\gsim 10^4$ GeV, that is $\alpha \geq 1.4$.

\section{Summary and Discussion}
\label{sec.6}

In this paper, we have studied quitessence models in a brane world
scenario.
We have adopted the second Randall-Sundrum brane scenario for a concrete
model, although a similar result would be obtained in other  brane world
models. As a consequence of a brane embedded in the
extra-dimension, the quadratic term of energy density appears,
changing the expansion law in   the
early stage of the universe. This
affects the dynamics of the scalar field in the quadratic-term dominant
stage. We have then investigated three candidates  for a
successful quintessence.

As a first candidate, we have discussed an inverse-power-law potential
model,
$V=\mu^{4+\alpha}\phi^{-\alpha} $ with
$2<\alpha <6$. We have shown the solution in which the density
parameter   of a scalar field   decreases as
$a^{-4(\alpha-2)/(\alpha+2)}$ in the $\rho^2$-dominant stage. This
feature provides us wider  initial conditions for a successful quintessence.
In fact, even if the universe is initially in a scalar-field dominant,
it eventually evolves into the radiation dominant era in the
$\rho^2$-dominant stage, which guarantees a successful nucleosynthesis in
the conventional universe stage.

Although  initial conditions could be arbitrary because the present
solution is an attractor, we may have a natural initial condition for some
specific origin of a potential such as a fermion condensation.
If this is the case,  equipartition of each energy density is more
likely. Assuming such an equipartition,  we have shown constraints
  in $\mu-m_5$ plane  for a natural and  successful
quintessence scenario. The allowed region gets
wider as $\alpha$ is larger, because for larger $\alpha$, the density
parameter becomes  smaller when the conventional cosmology is recovered.
We  conclude that in order to explain naturally the observational value of
the dark energy by the present   scenario,
$\alpha\geq 4$ is required. This constraint also restricts the value of
the 5-dimensional Planck mass, e.g. for
$\alpha=5$, $4.06 \times 10^{-14}m_4 \lsim  m_5 \lsim  2.78 \times
10^{-13}m_4$.

We have also discussed   an exponential potential model
$V=\mu^4\exp(-\lambda \phi/m_4)$, although by itself it
may not provide a successful quintessence scenario.
In  the five-dimensional brane scenario,
$\tilde{\lambda}=\lambda m_5/m_4$ is expected to be  order unity.
If that is the case,  the kinetic term of the scalar field becomes eventually
dominant for any initial conditions, and the density parameter of the scalar
field  decreases in the quadratic-term dominant stage. In this
case, however,
$\lambda$ becomes too large to explain the present scalar field
dominance. We may need unnatural modification in the potential.
  On the other hand, if $\lambda \sim O(1)$ to find a  successful
quintessence scenario in the conventional  universe stage,
$\tilde{\lambda}$ becomes very large. This provides an  extremely flat
potential, which  behaves just as a cosmological constant in
the $\rho^2$-dominant stage, resulting in an inflationary expansion
before reaching the conventional universe. After that, the radiation
never dominates, which contradicts nucleosynthesis.
Therefore, in both cases, we may not find a natural and successful
quintessence scenario.

As a third model, we have investigated a kinetic-term quintessence
(the so-called $k$-essence) model. We have adopted a model with
an inverse-power-law potential in coefficients of kinetic terms.
This  provides a tracking solution just the same as the
case with an inverse-
power-law potential.
We then  expect to obtain a natural quitessence scenario.
  Contrary to our
expectation, however, we do not find any solution in $\rho^2$-dominant
stage by which the density parameter $\Omega_\Phi$ decreases. Instead, we
find a tracking solution in which density parameter increases more slowly
than that in the conventional universe.
We also find a  scaling solution which is a transient attractor.
Then, if the universe starts with radiation dominance, the density
parameter keeps constant in the early stage and then the universe moves to
a tracking solution, finding a usual $k$-essence in the conventional
universe.
We do not find so much  advantage in a brane world.
Only the density parameter increases more slowly in the
$\rho^2$-dominant stage, which provides a wider initial condition for a
sucsessful quintessence.
Finally , we have shown the value of the
parameter $\mu$ appeared in this model can be taken as
the same order as the five-dimensional Planck mass scale if
$\alpha \geq 1.4$.

\section*{ACKNOWLEDGMENTS}
We would like to thank T. Harada, J. Koga and K. Yamamoto for useful
discussions and comments.  This work was  partially supported by the Waseda
University Grant for Special Research Projects and  by the Yamada
foundation.


\begin{thebibliography}{99}
\bibitem{CMB}
P. de Bernardis et al. , Nature {\bf404}, 955 (2000);  A. E. Lange et al.,
Phys.  Rev.  D {\bf 63},  042001 (2001); A. Balbi et al., Ap. J. {\bf545},
L1 (2000).
\bibitem{S-novae}
S. Perlmutter et al., Nature {\bf391}, 51 (1998); A. G. Riess et al.,
Astron. J. {\bf116},  1009 (1998); P. M. Garnavich et al., Ap. J.
{\bf509}, 74 (1998); S. Perlmutter et al. ,  Ap. J. {\bf517}, 565 (1999).
\bibitem{C. Problem}
Weinberg, Rev. Mod. Phys. {\bf61}, 1 (1989); V. Sahni and A. Starobinsky,
Int. J. Mod. Phys. D  {\bf9}, 373 (2000) ;
   S. Weinberg, astro-ph/0005265.
\bibitem{D-C-C}
A. D. Dolgov, in {\it The Very Early Universe} (eds. G. W. Gibbons et al.
,  Cambridge University Press), 449 (1982); L. H. Ford, Phys. Rev. D {\bf
35}, 2339 (1987); S. M. Barr, Phys. Rev. D {\bf36}, 1691 (1987); Y. Fujii
and T. Nishioka, Phys. Rev. D {\bf42}, 361(1990);  K. Coble, S. Dodelson
and J. A. Frieman, Phys. Rev. D {\bf55}, 1851 (1997).
\bibitem{Overduin}
J. M. Overduin and F. I. Cooperstock, Phys. Rev. D {\bf 58}, 043506
(1998).
\bibitem{Quintessence}
R. R. Caldwell, R. Dave and P. J. Steinhardt, Phys. Rev. Lett. {\bf 80},
1582 (1998); L. Wang, R. R. Caldwell, J. P. Ostriker and P. J. Steinhardt,
Astrophys. J. {\bf 530},17 (2000).

\bibitem{Ratra-Peebles}
B. Ratra and P. J. E. Peebles, Phys. Rev. D {\bf 37}, 3406 (1988) ; A.R.
Liddle  and R. J. Scherrer  Phys.   Rev.  D {\bf 59}, 023509 (1998).
\bibitem{tracker solutions}
I. Zlatev, L. Wang and P. J. Steinhardt, Phys. Rev. Lett. {\bf 82}, 896
(1999);  P. J. Steinhardt, L. Wang and I. Zlatev, Phys. Rev. D {\bf 59},
123504 (1999).
\bibitem{scaling  solutions}
P. G. Ferreira and M. Joyce, Phys. Rev. D {\bf58}, 023503 (1998);
E. J. Copeland, A. R. Liddle and
D. Wands, Phys. Rev. D {\bf57}, 4686 (1998); P. G. Ferreira and M. Joyce,
Phys. Rev. Lett. {\bf79},  4740 (1997); P. Viana and A. Liddle, Phys. Rev.
D {\bf57}, 674 (1998).
\bibitem{Grad}
Y. Fujii, Phys. Rev. {\bf 62}, 064004 (2000);
L. Amendola, Phys. Rev. D {\bf 62}, 043511 (2000);
A.Albrecht and C. Skordis,  Phys. Rev. Lett. {\bf 84},
2076 (2000);
S. Dodelson, M. Kaplinghat and E. Stewart, Phys. Rev. Lett. {\bf 85}, 5276
(2000).
\bibitem{Ex-exponential}
T. Barreiro, E. J. Copeland and N. J. Nunes, Phys. Rev. D 
{\bf 61}, 127301 (2000); 
V. Sahni and L. Wang, Phys. Rev. D {\bf 62}, 103517 (2000).
\bibitem{String}
N. Arkani-Hamed, S. Dimopoulos and G. Dvali,  Phys. Lett. B {\bf 429}, 263
(1998);  I. Antoniadis, N. Arkani-Hamed, S. Dimopoulos and G. Dvali,
Phys. Lett. B {\bf 436}, 257 (1998).
\bibitem{H-W}
P. Ho$\check{\rm r}$ava and E. Witten, Nucl. Phys. B {\bf 460}, 506 (1996);
ibid B {\bf 475}, 94 (1996).
\bibitem{RS}
L. Randall and R. Sundrum, Phys. Rev. Lett. {\bf 83},  4690 (1999);
L. Randall and R. Sundrum, Phys. Rev. Lett. {\bf 83},  3370 (1999).
\bibitem{3-brane}
V. A. Rubakov and M. E. Shaposhinikov, Phys. Lett. B {\bf 125}, 139
(1983); K. Akama, in  {\it Gauge Theory and Gravitation} ed by K. Kikkawa,
N. Nakanishi, and H. Nariai  (Springer-Verlag, 1983); K. Akama,
hep-th/0001113.
\bibitem{SMS}
T.  Shiromizu, K.  Maeda, and M.  Sasaki, Phys.  Rev.  D {\bf 62},
024012 (2000).
\bibitem{Brev}
R. Maartens, gr-qc/0101059.
\bibitem{Bcmgy1}
P. Bin$\acute{e}$truy, C. Deffayet and D. Langlois, Nucl. Phys. B  {\bf
565}, 269 (2000); N. Kaloper, Phys.  Rev.  D {\bf 60}, 123506 (1999) ; C.
Csaki, M. Graesser, C. Kolda  and J. Terning Phys. Lett. B {\bf 462}, 34
(1999) ; T. Nihei, Phys. Lett. B {\bf 465},  81 (1999) ; P. Kanti, I. I.
Kogan, K. A. Olive and M. Prospelov, Phys. Lett. B {\bf 468}, 31 (1999);
J. M. Cline, C. Grojean and G. Servant, Phys. Rev. Lett. {\bf 83},  4245
(1999);  P. Bin$\acute{e}$truy, C. Deffayet, U. Ellwanger and D. Langlois,
Phys. Lett. B {\bf 477}, 285 (2000) ;  S. Mukohyama, T. Shiromizu and K.
Maeda, Phys. Rev. D {\bf 62}, 024028 (2000).
\bibitem{Bcmgy2}
R. Maatens, D. Wands, B. A. Bassett and I. P. C. Heard, Phys. Rev. D  {\bf
62}, 041301 (2001); L. Mendes and A. R. Liddle, Phys. Rev. D {\bf 62},
103511 (2000);  J. Khoury, P. J. Steinhardt and D. Waldram, Phys. Rev. D
{\bf 63}, 103505 (2001);  A. Mazumdar Nucl. Phys. B {\bf 597}, 561 (2001);
R. M. Hawkins and J. E. Lidsey, Phys. Rev. D  {\bf 63}, 041301 (2001);
S. Tsujikawa, K. Maeda and S. Mizuno, Phys. Rev. D, {\bf 63}
   ,123511 (2001); S. C. Davis, W. B. Perkins, A.-C. Davis and I.R. Vernon,
Phys. Rev. D {\bf 63},  083518 (2001).
\bibitem{Bcmgy3}
E. J. Copeland, A. R. Liddle and J. E. Lidsey, astro-ph/0006421
; G. Huey and J. E. Lidsey, astro-ph/0104006.
\bibitem{maeda}
K. Maeda, astro-ph/0012313.
\bibitem{maeda_wands}
K. Maeda, and D. Wands, Phys. Rev. D {\bf 62},  124009 (2000).
\bibitem{ivplpotentilorigin}
   P. Bin$\acute{e}$truy, Phys. Rev. D {\bf 60}, 063502 (1999); P. Brax  and
J. Martin, Phys.   Rev.  D {\bf 61}, 103502 (2000); T. R. Taylor, G.
Veneziano and
   S. Yankielowicz, Nucl. Phys. B {\bf 218}, 493 (1983); I. Affleck,  M.
Dine and N. Seiberg,
   Nucl. Phys. B {\bf 256}, 557 (1985).
\bibitem{Future}
K. Maeda, S. Mizuno and K.Yamamoto, in preparation.
\bibitem{expotentilorigin}
B. Whitt, Phys. Lett. B {\bf145}, 176 (1984);
J. D. Barrow and S. Cotsakis,  Phys. Lett. B {\bf214}, 515 (1988);
D. Wands, Class. Quantum Grav. {\bf11},269 (1994);
M. B. Green, J. H. Schwarz, and E. Witten, {\it Superstring Theory}
( Cambridge University Press, Cambridge, England, 1987).
\bibitem{K-Quintessence}
T. Chiba, T. Okabe and M. Yamaguchi, Phys. Rev. D {\bf 62}, 023511.
\bibitem{K-essence}
C. Armendariz-Picon, V. Mukhanov and P. J. Steinhardt, Phys. Rev. Lett.
{\bf 85}, 4438 (2000).
\end{thebibliography}
\end{document}